\documentclass[fleqn,usenatbib]{mnras}
\usepackage{newtxtext,newtxmath}

\usepackage{lscape}
\usepackage{graphicx}
\usepackage[space]{grffile}
\usepackage{latexsym}
\usepackage{amsfonts,amsmath,amssymb}
\usepackage{siunitx}
\usepackage{url}
\usepackage[utf8]{inputenc}
\usepackage{hyperref}
\hypersetup{pdfborder={0 0 0}}
\usepackage{adjustbox}
\usepackage{textcomp}
\usepackage{longtable}
\usepackage{mathrsfs}
\usepackage{multirow,booktabs}
\usepackage{upgreek}
\usepackage{threeparttable}
\usepackage{tablefootnote}
\usepackage{caption}
\usepackage{color}
\usepackage{subfig}
\usepackage{cleveref}

\usepackage[T1]{fontenc} 

\crefformat{section}{\S#2#1#3} 
\crefformat{subsection}{\S#2#1#3}
\crefformat{subsubsection}{\S#2#1#3}

\DeclareMathSymbol{\ast}{\mathbin}{symbols}{"03}

\title[UTMOST real-time FRB survey]{Five new real-time detections of Fast Radio Bursts with UTMOST}

\author[Farah, W. et al.]{\parbox{\textwidth}
   {W.\ Farah$^{1}$, 
    C.\ Flynn$^{1,2}$,
    M.\ Bailes$^{1,3}$, 
    A.\ Jameson$^{1,2}$, 
    T.\ Bateman$^{4}$, 
    D.\ Campbell-Wilson$^{4}$, 
    C.\ K.\ Day$^{1}$,
    A.\ T.\ Deller$^{1,2}$, 
    A.\ J.\ Green$^{4}$, 
    V.\ Gupta$^{1}$, 
    R.\ Hunstead$^{4}$, 
    M.\ E.\ Lower$^{1,5}$,  
    S.\ Os{\l}owski$^{1}$, 
    A.\ Parthasarathy$^{1}$,
    D.\ C.\ Price$^{1,6}$,
    V.\ Ravi$^{7,8}$, 
    R.\ M.\ Shannon$^{1,3}$, 
    A.\ Sutherland$^{4}$, 
    D.\ Temby$^{4}$,
    V.\ Venkatraman Krishnan$^{1,2,9}$, 
    M.\ Caleb$^{10,11}$, 
    S.-W.\ Chang$^{2,3,11}$, 
    M.\ Cruces$^{9}$,
    J.\ Roy$^{12}$, 
    V.\ Morello$^{10}$,
    C.\ A.\ Onken$^{2,3,11}$,
    B.\ W.\ Stappers$^{10}$,
    C.\ Wolf$^{2,11}$}  \\ \\ \\
\parbox{\textwidth}{
$^1$Centre for Astrophysics and Supercomputing,
  Swinburne University of Technology, Mail H30, PO Box 218, VIC 3122,
  Australia\\
$^{2}$ARC Centre of Excellence for All-sky Astrophysics (CAASTRO)\\
$^{3}$ARC Centre of Excellence for Gravitational Wave Discovery (OzGrav), Australia\\
$^{4}$Sydney Institute for Astronomy, School of Physics A28, University
  of Sydney, NSW 2006, Australia\\
$^{5}$CSIRO Astronomy and Space Science, Australia Telescope National Facility, Epping, NSW 1710, Australia\\
$^{6}$Department of Astronomy, University of California Berkeley, 501 Campbell Hall, Berkeley CA 94720\\
$^{7}$Center for Astrophysics | Harvard \& Smithsonian, 60 Garden Street, Cambridge MA 02138, USA\\
$^{8}$Cahill Centre for Astronomy and Astrophysics, MC 249-17, California Institute of Technology, Pasadena, CA 91125, USA\\
$^{9}$Max-Plank-Institute f\"{u}r Radioastronomie, Auf dem H\"{u}gel 69, D-53121 Bonn, Germany\\
$^{10}$Jodrell Bank Centre for Astrophysics, School of Physics and Astronomy, The University of Manchester, Manchester M13 9PL, UK\\
$^{11}$Research School of Astronomy and Astrophysics, Australian National University, Canberra, ACT 2611, Australia\\
$^{12}$National Centre for Radio Astrophysics, Tata Institute of Fundamental Research, Pune 411 007, India
}
}

\newcommand{\dm}{$\textrm{pc}\,\textrm{cm}^{-3}$}

\date{Accepted XXX. Received YYY; in original form ZZZ}

\pubyear{2019}

\begin{document}
\label{firstpage}
\pagerange{\pageref{firstpage}--\pageref{lastpage}}
\maketitle

\begin{abstract}
We detail a new fast radio burst (FRB) survey with the Molonglo Radio Telescope, in which six FRBs were detected between June 2017 and December 2018. By using a real-time FRB detection system, we captured raw voltages for five of the six events, which allowed for coherent dedispersion and very high time resolution (10.24 $\upmu$s) studies of the bursts. Five of the FRBs show temporal broadening consistent with interstellar and/or intergalactic scattering, with scattering timescales ranging from 0.16 to 29.1 ms. One burst, FRB181017, shows remarkable temporal structure, with 3 peaks each separated by 1 ms. 
We searched for phase-coherence between the leading and trailing peaks and found none, ruling out lensing scenarios. Based on this survey, we calculate an all-sky rate at 843 MHz of $98^{+59}_{-39}$ events sky$^{-1}$ day$^{-1}$ to a fluence limit of 8 Jy-ms: a factor of 7 below the rates estimated from the Parkes and ASKAP telescopes at  1.4 GHz assuming the ASKAP-derived spectral index $\alpha=-1.6$ ($F_{\nu}\propto\nu^{\alpha}$). Our results suggest that FRB spectra may turn over below 1 GHz. Optical, radio and X-ray followup has been made for most of the reported bursts, with no associated transients found. No repeat bursts were found in the survey. \\

\end{abstract}

\section{Introduction}\label{sec:introduction}
Even though more than a decade has passed since they were first detected, fast radio bursts (FRBs) 
still defy explanation. Discovered by \cite{Lorimer2007}, FRBs are millisecond-wide bursts seen in the 
radio part of the electromagnetic spectrum. The observed integrated electron column density, i.e. dispersion measure (DM), 
along the lines of sight of FRBs significantly exceeds that expected 
from the Milky Way, placing FRB sources sources at cosmological distances if the intergalactic medium 
(IGM) is the major contributor to the excess DM \citep{Shannon2018}.

Of the 69 FRBs published to date (FRBCAT\footnote{\href{http://frbcat.org}{http://frbcat.org}; visited 11/04/2019}; \citealt{FRBCat}), 
only two have been seen to repeat. The repeat bursts of FRB121102 allowed for 
an unambiguous localisation of the FRB source which resides in a star-forming region of a dwarf galaxy at redshift $z=0.193$
\citep{Chatterjee2017,Marcote2017,Bassa2017,Tendulkar2017}. A large Rotation Measure (RM) of 
$10^5$\,rad\,m$^{-2}$ reported by \cite{Michilli2018} places this FRB source in an extreme magneto-ionic environment. 
With the more recently discovered repeater FRB180814.J0422+73 by the CHIME radio telescope \citep{Amiri2019_repeater}, repeating FRBs seem to share 
common characteristics, namely pulse-to-pulse variation with bursts showing complex temporal and spectral structure \citep{Hessels2018}.
A few non-repeating FRBs show similar structure (e.g. \citealt{Farah2018,Ravi2016_science}). This appears 
to be the only bridge connecting the potentially bifurcated classes, given that they occupy different regions of phase-space 
\citep{Palaniswamy2017}, and that non-repeaters show modest RM (\citealt{Caleb2018}; Os{\l}owski et al. in prep.).
Sub-pulse frequency drifts seen in the repeating FRBs are reminiscent of solar type III radio bursts, suggesting an analogous 
emission mechanism \citep{Amiri2019_repeater}.

Scattering is characteristic of a pulsed radio signal traversing turbulent media, where the delayed 
time of arrival due to multipath propagation is manifested as an exponential tail in the signal pulse profile. 
It is not surprising that FRBs are under-scattered with respect to 
Galactic pulsars with the same DM \citep{Ravi2019_observed_prop}, 
given that the bulk of the FRB DM is likely to be due to propagation through the IGM \citep{Shannon2018}, 
which is thought to be less turbulent and hence less effective at scattering radio 
waves compared to the ISM \citep{Koay&Macquart2015}. 
However, evidence supporting the existence of a scattering timescale $\tau$-DM 
relation for FRBs is accumulating \citep{Amiri2019_13frbs,Ravi2019_observed_prop}, 
suggesting that scattering takes 
place in the IGM, possibly in the circumgalactic gas clumps of intervening galaxies \citep{Vedantham2019}.
The scattered rays of radio emission of FRBs can also interfere with each other, 
giving rise to diffractive scintillation, evident as spectral modulation in the dynamic spectra of FRBs (e.g. \citealt{Masui2015,Ravi2016_science,Farah2018}). Plasma lensing 
arising from scattering regions can enhance the radio flux of 
FRBs \citep{Main2018_plasma_lensing} or even produce multiple images of the same burst with arrival times a few ms apart \citep{Cordes2017}.

Given their inferred cosmological distances, FRBs offer a means to probe the baryonic content of the IGM 
\citep{Deng&Zhang2014,Munoz2018,Ravi2019_whitepaper} and galaxy halos \citep{McQuinn2014}. Moreover, FRBs can 
also probe the existence of massive compact halo objects (MACHOs) if such objects are fortuitously aligned with FRB lines of sight \citep{Zheng2014}. The strong gravitational lensing of 
an FRB by a MACHO in the mass range of $20$-$100\ \rm M_{\odot}$ 
would result in multiple images of the burst \citep{munoz2016_lensing}. 
Although the images would appear at an angular separation well below the resolving power of radio telescopes, 
the time of arrival of the 
pulses will differ by a few $\times (\rm M_L/30\rm M_{\odot})$ ms, 
where $ \rm M_L$ is the mass of the lens.
Only if phase information is available, phase coherence can be searched for in 
temporarily-resolved multi-peaked FRBs in order to test lensing scenarios.

New generation telescopes are promising to revolutionise the FRB field in the very near future. ASKAP \citep{Shannon2018} and CHIME \citep{Amiri2019_13frbs} nearly doubled the total number of known FRBs only in the last year. 
The real-time FRB discovery system recently deployed on ASKAP 
will allow voltage capture that, in turn, can be used to image the sky, delivering a host galaxy association.
The large ($\sim250$ deg$^2$) field of view of CHIME will allow the discovery of FRBs at a rate of a few per day \citep{Connor2016_rates}. The Molonglo Observatory Synthesis Telescope (MOST) has been undergoing a transformation into an FRB-finding machine \citep{utmost}. \cite{Caleb_3frbs} reported the discovery of 
the first FRBs using this interferometer, placing the FRB source at least $>10^4$ km away from the telescope. 
More recently, \cite{Farah2018} reported the blind detection of FRB170827 where the phase information of the 
detected radiation was preserved in the recorded data owing to its real-time discovery. 
Detailed analysis of the coherently dedispersed data of FRB170827 revealed rich spectral and temporal structure.
UTMOST-2D is a project currently underway to fit the North-South (NS) arms of the Molonglo radio telescope 
with outriggers and a central detector to 
achieve arcsecond localisation of FRBs (Day et al. in prep.). Other surveys dedicated to FRB searches 
are also currently in progress or in development 
\citep{vfastr_wayth2011,apertif,meertrap_stappers2018,SUPERB,Realfast,OWFA_GMRT_Bhattacharyya2018,greenburst_surnis2019}. 
It is becoming standard to make use of machine learning algorithms 
to perform FRB candidate classification. 
Different approaches have been taken by different groups. For example, 
the FRB discovery pipelines described by \cite{Wagstaff2016} and 
\cite{Foster2018} are based on the traditional probabilistic machine learning algorithm random forest.
Conversely, deep learning is also emerging as a promising technique 
for FRB discovery \citep{Connor2018,Zhang2018_ML,Agarwal2019_ML}.

In this paper, we report the discovery of five new fast radio bursts using the Molonglo radio telescope. 
We summarise the observing set-up and time-on-sky spent searching for FRBs in \cref{sec:observation}. 
In \cref{sec:pipeline}, we describe our machine-learning based, real-time FRB detection pipeline. 
We detail our new discoveries in \cref{sec:frb_discoveries}, and derive our FRB rates in \cref{sec:rates}. 
We describe the follow-up campaign in \cref{sec:frb_followup} and draw our conclusions in \cref{sec:conclusions}.

\section{UTMOST and FRB searches}\label{sec:observation}


MOST is located some 40 km east of Canberra, Australia. It
is a Mills-Cross interferometer, comprised of two fully steerable east-west (EW) arms, each 778m long 
with a total of 18000 m$^2$ collecting area. 
The UTMOST project transformed the MOST into a commensal pulsar-timing/FRB-finding facility \citep{utmost}, 
operating at 843 MHz, with a bandwidth of 31.25 MHz. 
Using this telescope, nine FRBs have been found to date. 
Three of these are reported in \cite{Caleb_3frbs}, and another is reported in detail in \cite{Farah2018}. 
In this paper, we describe the five additional events in detail and derive improved population properties of FRBs at 
843 MHz.

\cite{Caleb_3frbs} estimated a rate of 78$^{+124}_{-57}$ events sky$^{-1}$ day$^{-1}$ at 843 MHz 
above a fluence of 11 Jy-ms (a limit we revise to 15 Jy-ms, see \cref{sec:rates}). These first 
three FRBs were found when the system had frequency channels 0.78 MHz-width so the effects of DM smearing 
were quite pronounced. The system has since been upgraded to 0.097 MHz-width channels, significantly improving our spectral 
resolution for the subsequent FRBs. The temporal resolution has been also improved 
from 655 $\upmu$s to 327 $\upmu$s, increasing our sensitivity to events narrow in time.

To search for FRBs, Molonglo’s $4 \times 2.8 $ square degree primary beam is tiled with consecutive,
overlapping narrow strips. These ``fan-beams'' are narrow in the EW direction (full width half maximum (FWHM) $\approx 
45''$), but broad in the north south direction (FWHM $\approx 2.8^{\circ}$), meaning that host galaxy identification is not possible for detected FRBs. UTMOST-2D, a project currently under development, will make use of the NS arms of the telescope to achieve arcsecond localisation of FRBs.

\subsection{Live FRB discovery pipeline}
The telescope operates in a band affected by interference 
caused by mobile phone transmissions from handsets. 
These sources of radio frequency interference (RFI) dominate false positives and were typically
removed via human inspection of the data each morning.
We describe here a fully automated system that performs this classification on the live data 
sufficiently rapidly to achieve voltage capture of the data for good candidates. 

Voltage capture of interesting events is made in narrow time windows that encompass the dispersion smearing time, 
taking place after a real-time detection and classification
before the observations are down-sampled and saved to disk. 
The time and frequency resolutions of UTMOST’s final 
data product for human inspection after voltage capture are, respectively, 8 and 64 times 
higher than the data retained for usual offline analysis.
The FRBs detected by \cite{Caleb_3frbs} using the offline pipeline are sampled at 655 $\upmu$s and 0.78 MHz; 
structure on smaller time and frequency intervals was completely unseen in the data.


Moreover, search-mode data suffer from interchannel DM-smearing, and algorithms usually reverse the effect of dispersion 
by shifting each individual channel backwards in time --- a process called incoherent dedispersion.
On the other hand, coherent dedispersion makes use of the phase information preserved in raw data 
(complex voltages) of the receiver in order to completely correct for dispersion.
However, this process is computationally expensive and is rarely used when searching blindly  
for FRBs in real time.  
 
\subsection{Sensitivity improvements}

The sensitivity of the EW arms was substantially improved in 2017 after converting
the facility into a transit-only instrument only. 
Although the advantage of UTMOST's rotating ring antennas was achieving mechanical 
phasing in the EW direction, breakages and faults occurred on regular basis, and, 
thus, the EW slewing system was retired.

The 7744 ring antennas were aligned to the meridian over a four month period from early-to mid-2017. 
This was performed on a module-by-module basis, and regular observations of the bright pulsar Vela 
transiting the meridian were performed to validate the alignment and track the sensitivity increases. 
The result was a factor $\approx$ 2 increase on average in the system sensitivity, which was achieved by June 2017. 
Since then, observations have been done entirely in transit mode, as the object of interest crossed the meridian. 
\subsection{Time on sky}

Observations at MOST are performed almost completely autonomously using the 
dedicated Survey for Magnetars, Intermittent pulsars, RRATs and FRBs (SMIRF) scheduler. 
While the comprehensive description of the software is left to an upcoming paper (Venkatraman Krishnan et al., in prep.), 
we briefly describe its mode of operation. SMIRF schedules which 
fields to observe, given local sidereal time and a pre-defined cadence list of FRB fields, pulsars, and pulsar 
search-pointings. A unique feature of UTMOST and SMIRF is that pulsar timing, periodicity and single-pulse pulsar 
searching, and FRB blind searching can be done commensally and in real time. 
This automated scheduler achieved very substantial efficiency gains over its precursor, 
in addition to the increased sensitivity, such that we can now 
regularly time about 400 pulsars on a weekly basis, do follow-up monitoring of known FRB fields and monitor the system sensitivity.
Moreover, the SMIRF scheduler has the 
potential to observe phase calibrators if needed, although this feature has yet to be used; human 
intervention is still necessary to decide on the quality of a calibration and whether or not a phase 
solution should be applied. In general, the system is proving to be stable enough that phase calibration need only 
be performed every few days, unless the phase solution is lost (e.g. to power outages).

\begin{figure}
  \begin{center}
    \includegraphics[width=\linewidth]{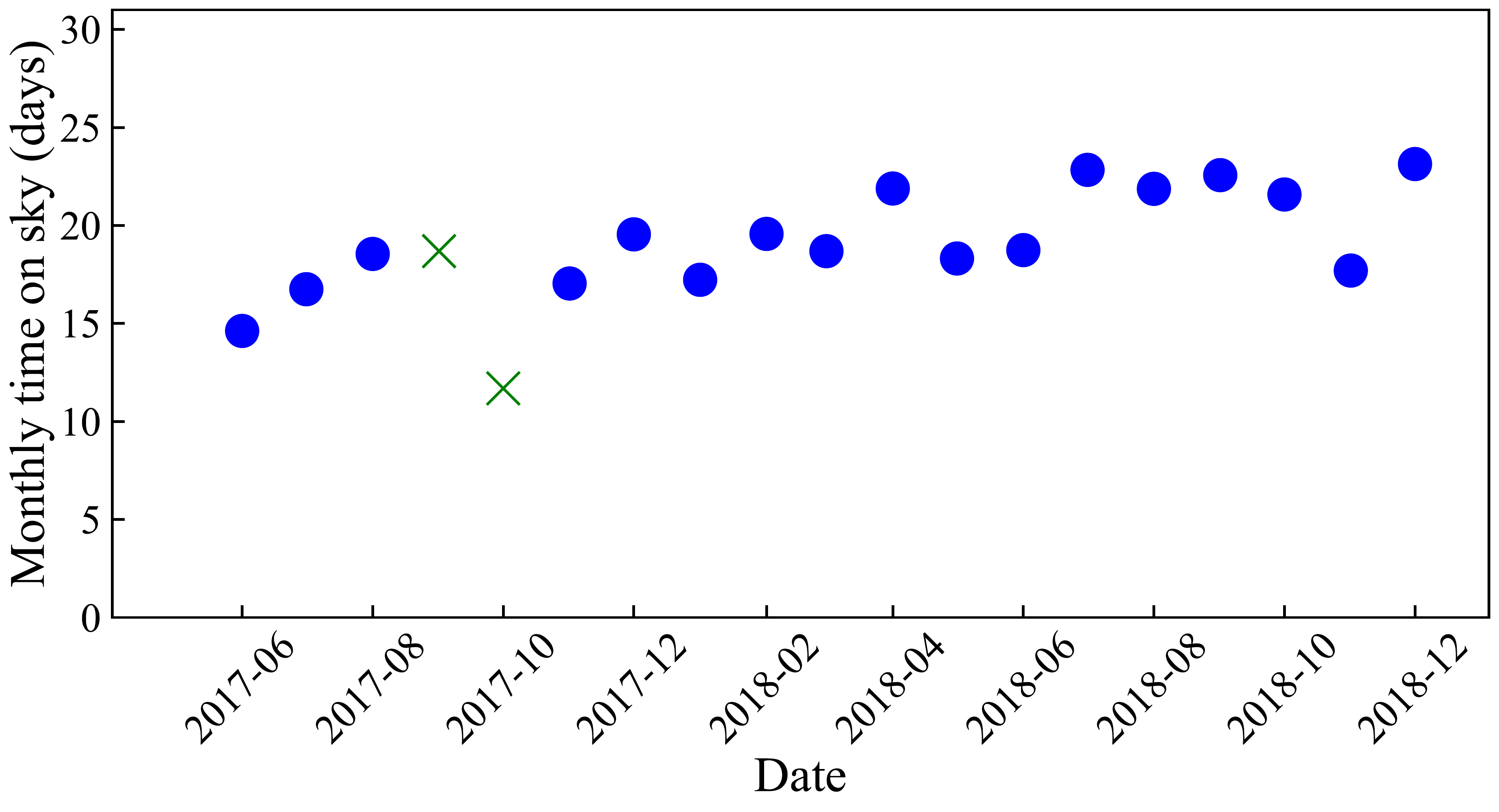}
    \caption{Number of days in each month of FRB-search time on sky over the course of the present survey. Blue circles show the monthly time-on-sky, while green crosses show two months where the time-on-sky had to be interpolated after a RAID failure led to the corruption of some metadata (search data were ordinarily processed prior to the failure). The efficiency of the system 
    has been steadily increasing since the deployment of the SMIRF scheduler, 
    along with the stability of the mechanical and electronic system.}
    \label{figure:time_on_sky}
  \end{center}
\end{figure}

\begin{figure}
    \centering
    \includegraphics[width=\columnwidth]{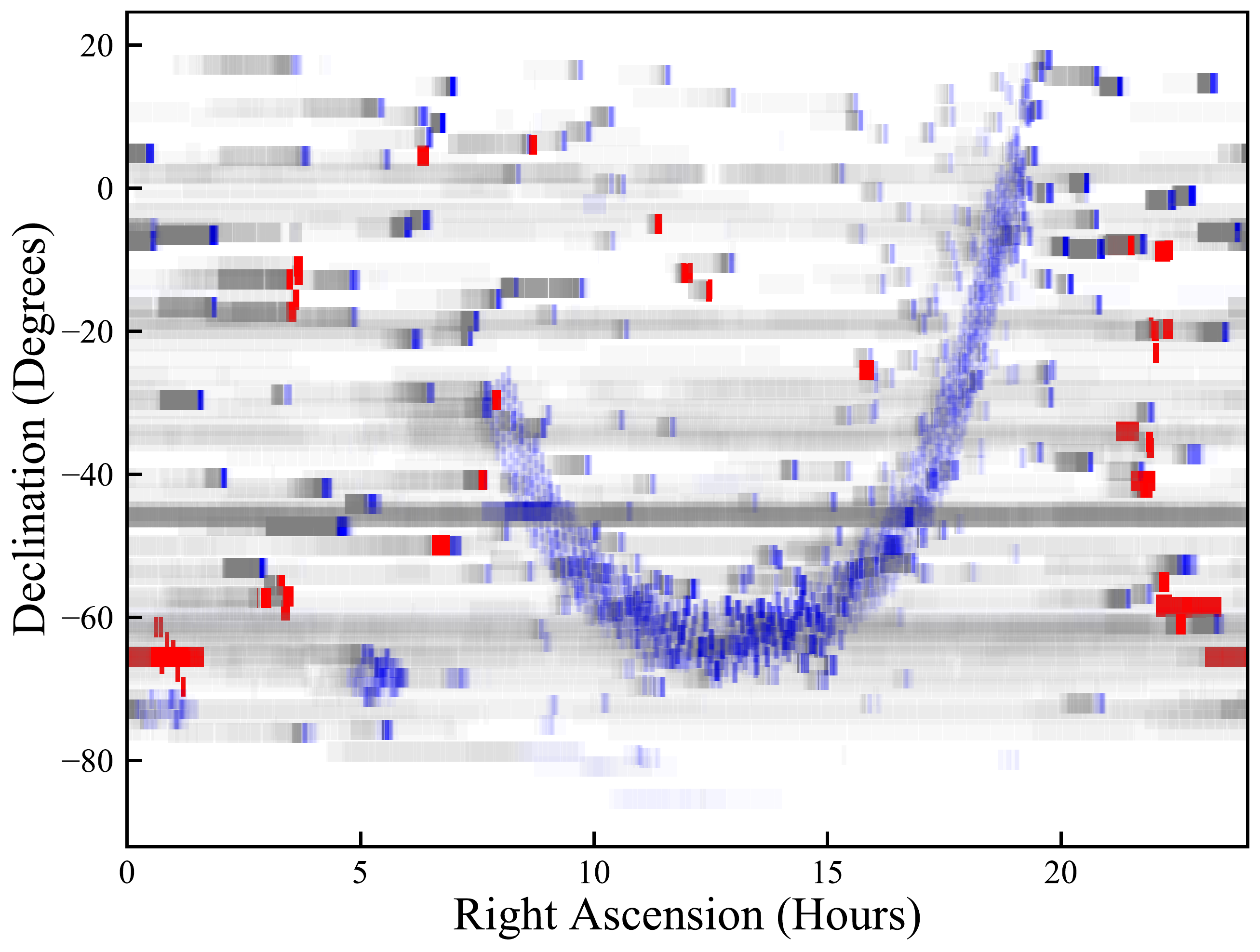}
    \caption{Regions of the sky surveyed by UTMOST in the time period between June 2017 and December 2018. 
      Grey represents observations of FRB-only fields, blue represents commensal pulsar observations/searches and FRB searches, and red regions mark 
      FRB fields followed up by UTMOST. Colour depth indicates the integration times on sky.}
    \label{fig:sky_survey}
\end{figure}

After the completion of the meridian drive and alignment of the EW feed antennas, 344 days on sky of 
FRB searching were completed between early June 2017 and December 2018. Fig.\,\ref{figure:time_on_sky} 
shows the monthly time on sky for the survey described above. A disk failure due to a power outage 
in October 2017 resulted in the corruption of meta-data for the months of September and October 2017. 
We replaced the corresponding 2 data points in Fig.\,\ref{figure:time_on_sky} 
for these months with the median of the monthly 
time on sky and median$-7$ days (to reflect the time lost on sky), respectively.
Fig.\,\ref{fig:sky_survey} shows in Right Ascension and Declination (RA, Dec) 
fields in which pulsars are timed or searched for in blue, fields in which we have done 
FRB follow-up in red, and finally grey shows fields where we 
solely search for FRBs, including 24-hour scans of the sky at fixed 
declination. This strategy is employed if one of the telescope arms fails, 
and over the summer break when no staff are on site. 
Our off-sky time is due to scheduled monthly maintenance, telescope repairs, 
slew time, calibration and weather conditions.

\section{FRB detection pipeline}\label{sec:pipeline}
UTMOST's real-time FRB discovery system is based on the graphics processing unit (GPU) program \textsc{heimdall} \citep{Barsdell_PhD}. 
\textsc{heimdall} performs dedispersion over a range of DM trials\footnote{increased to 5000 since October 2018; see text}
(0~-~2000 
$\textrm{pc}~\textrm{cm}^{-3}$) and then performs a variable width boxcar convolution on the timeseries 
to determine the optimal width of a candidate burst. Due 
to the harsh radio frequency interference (RFI) environment on site, \textsc{heimdall} 
produces candidates on the order of millions per day, with most being characterised 
as 5 MHz and a few millisecond-wide impulsive bursts. 
In order to deal with the large influx of candidates, we have developed a low 
latency machine-learning based candidate classification pipeline using the random forest algorithm
\citep{Breiman2001}. A random forest is a supervised machine-learning algorithm that can 
be described as an aggregation of multiple decision trees that, 
collectively, form a robust classifier or regressor. The classification system is described 
in detail in the following sections.

\begin{figure}
  \begin{center}
    \includegraphics[scale=0.3]{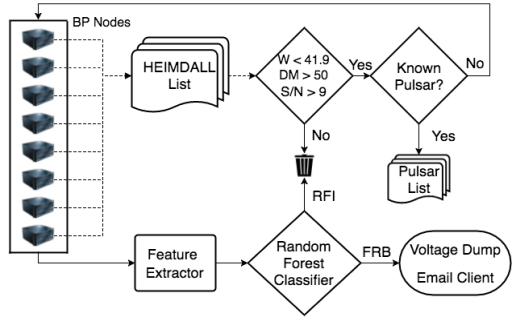}
    \caption{Schematic showing the signal path of the UTMOST detection pipeline.
      The processing time of any given candidate is typically $\sim 20$ seconds.}
    \label{figure:FRB_detector}
  \end{center}
\end{figure}

In Fig.\,\ref{figure:FRB_detector}, we show a schematic describing the signal path. 
Beamformed data (i.e. fan-beams) are analysed on the beam processing (BP) nodes by 
\textsc{heimdall}, where they are held in RAM typically for 24 seconds (for a detailed 
description of the UTMOST processing backend, see \cite{utmost}). 
The \textsc{heimdall} list of candidates is then checked against a known-pulsars 
list on a server. The list is then passed back to the respective BP node where 
feature extraction and candidate classification is performed. 
In order to successfully trigger a voltage capture, the runtime of the whole process 
should not exceed the length of the data on the RAM ring-buffers.

\subsection{Training set}

In general, a supervised machine-learning algorithm undergoes a phase of ``training'', where the 
algorithm is typically presented with a set of labelled data. The hyperparameters 
of the model are adjusted during the training phase 
such that the model is able to classify a similar but unfamiliar set as accurately as possible. 
A total of $\sim$10,000 candidates --- comprised of single pulses from various pulsars, artefacts, and RFI-contaminated data ---
were collected in order to build a 2-class training set used for the UTMOST real time classifier. 

\subsection{Pre-classifier candidate filtering}
A first stage of filtering is applied on the candidates output, from \textsc{heimdall}. All candidates with 
S/N $<9$, width $\geq 42~\textrm{ms}$, and DM $<50 \textrm{ pc\,cm}^{-3}$ are rejected as probable
artefacts. Each of the remaining candidates are then checked against a pulsar catalog and is 
marked as a from pulsar if its DM lies within 50\% of the pulsar's DM and its position on sky 
is within $\pm\,2$ fan-beams of the pulsar's position (a pulse has a chance to be detected simultaneously
in two neighbouring fan-beams, as the fan-beams are spaced a full-width-half-maximum apart in normal observing). 
Single pulses from pulsars are still presented to the classifier and logged; however, observers are not 
notified about these events. 

\begin{figure*}
    \centering
    \includegraphics[width=0.32\textwidth]{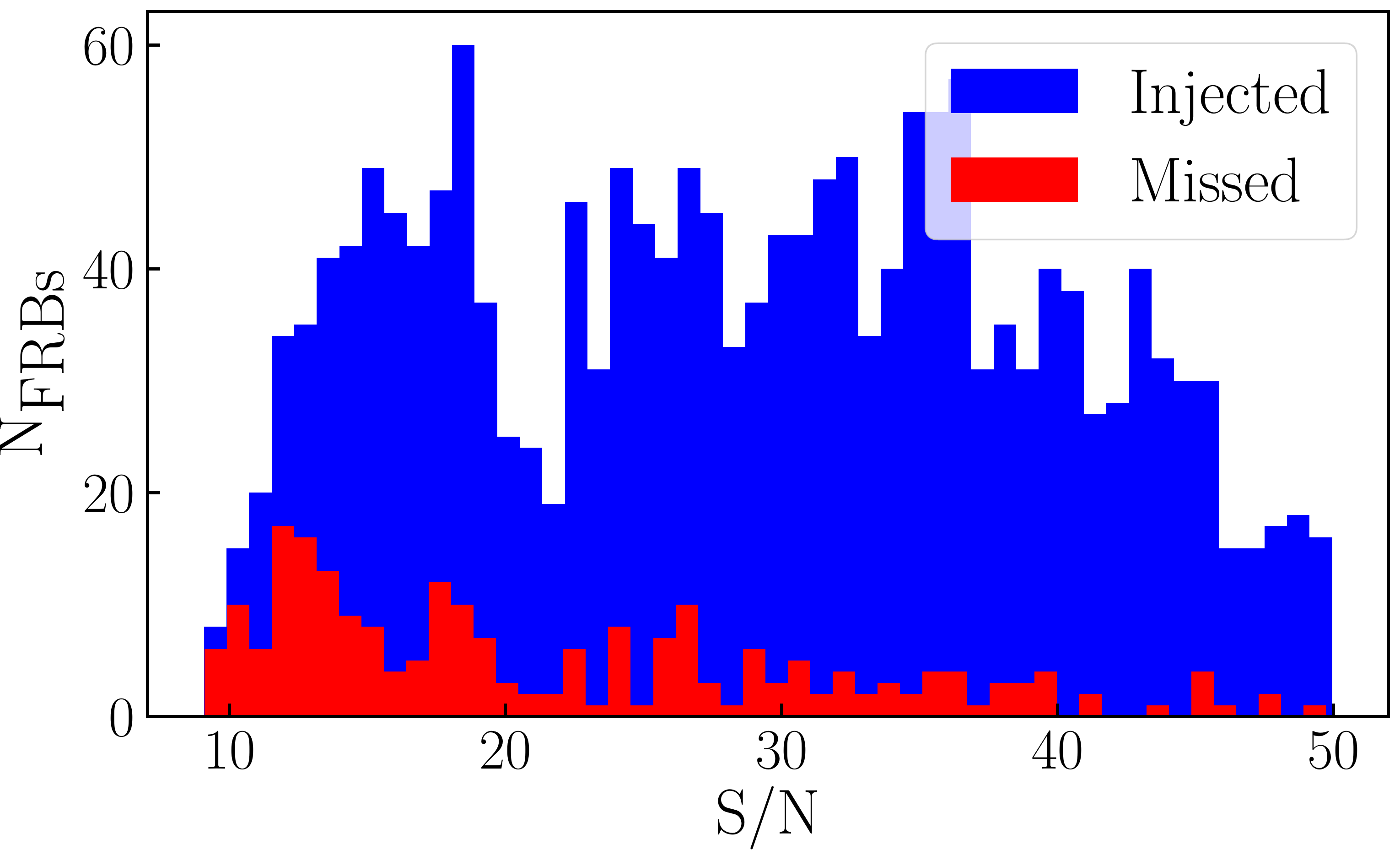}
    \includegraphics[width=0.32\textwidth]{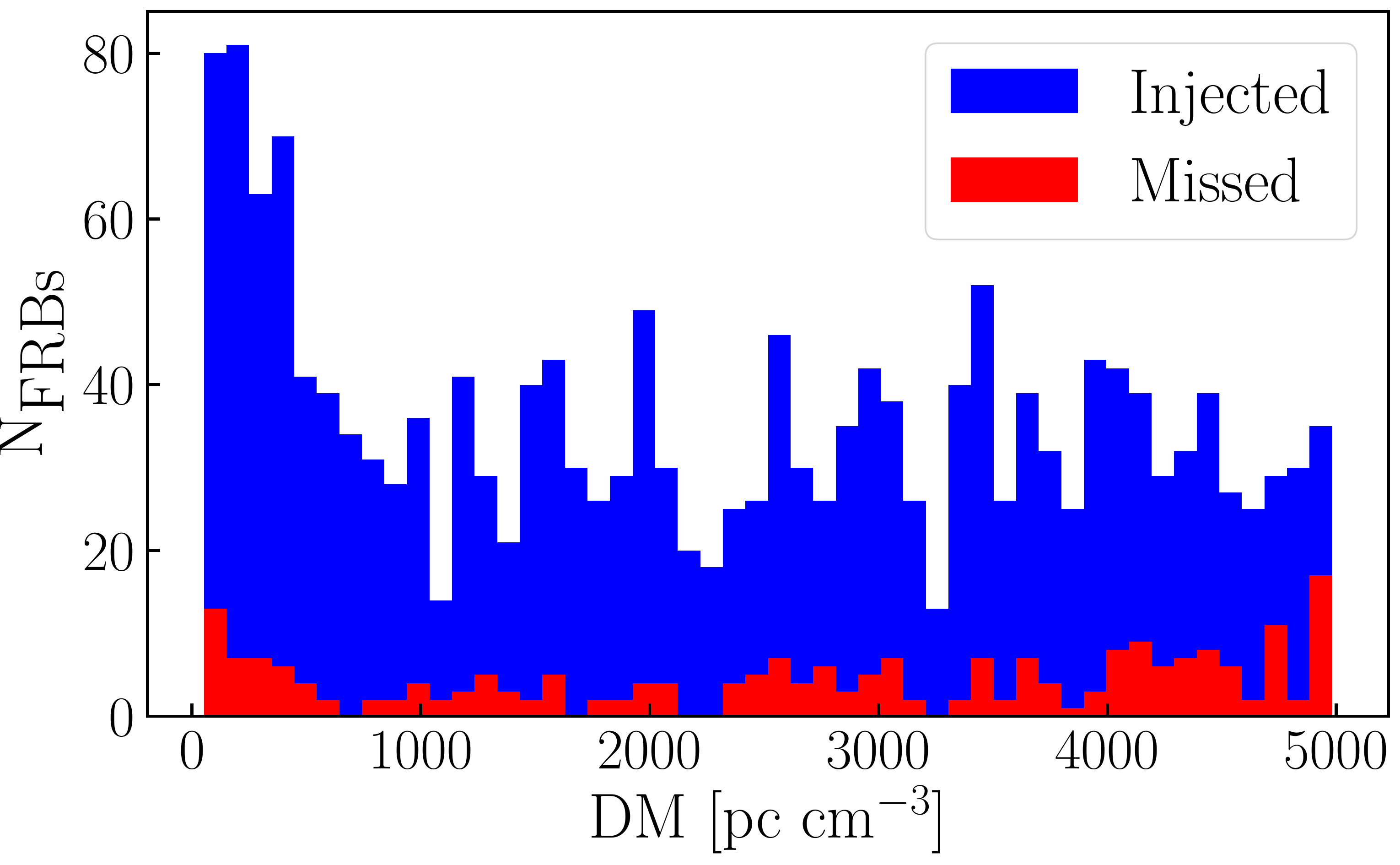}
    \includegraphics[width=0.32\textwidth]{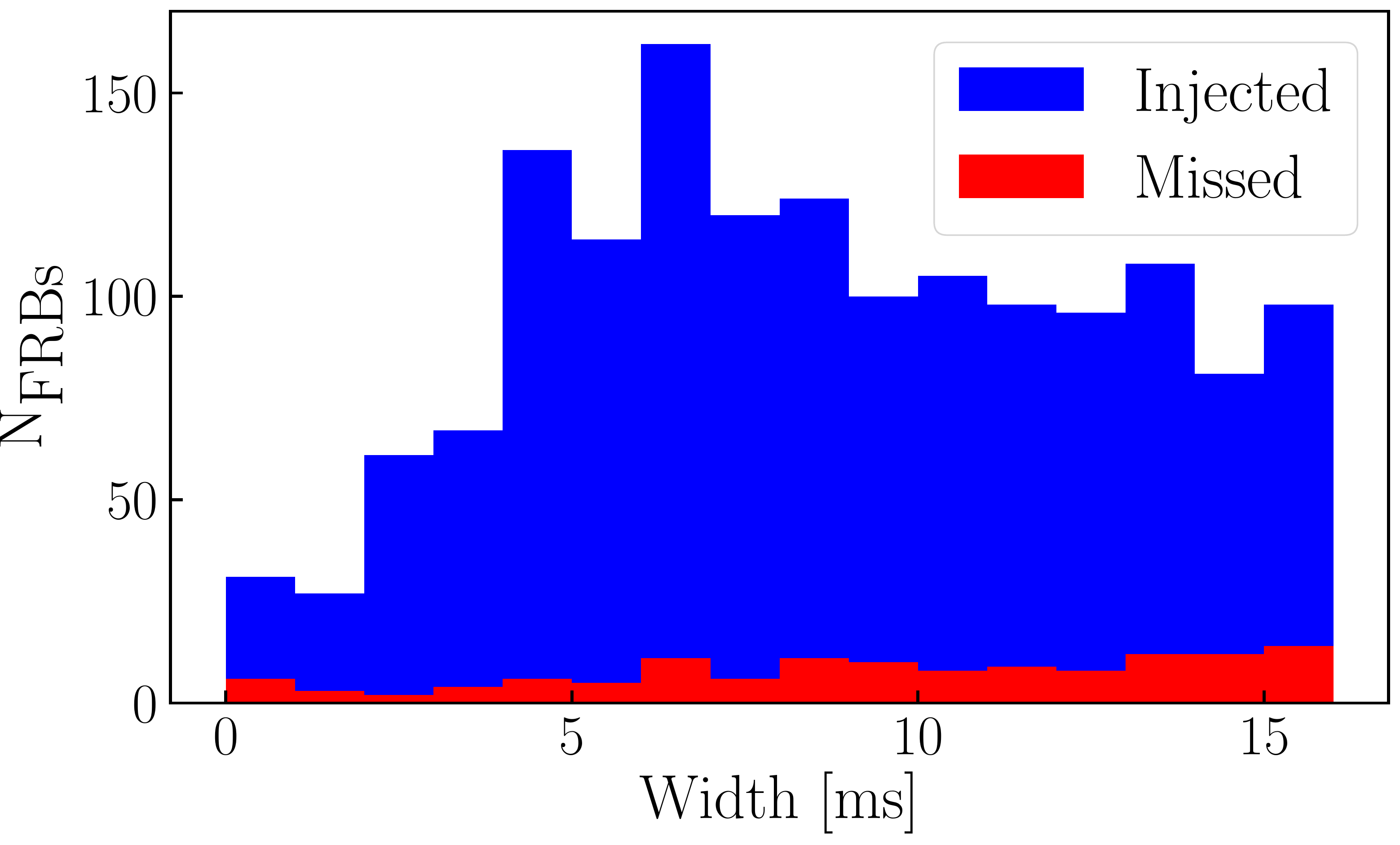}
    \caption{Distribution of S/N (left panel) DM (middle panel) and width (right panel) of the 
      $\sim 2000$ FRBs that were injected into UTMOST live stream data. The distribution of FRBs missed by our 
      pipelines are plotted in red.}
    \label{fig:injections}
\end{figure*}

\subsection{Feature extraction}

The candidates that pass the pre-classifier filter are input to a feature extraction stage, 
where a list of predictors are extracted from the frequency-time data. 
These features are carefully engineered statistics that are capable of characterising the noise and 
signal of a given candidate. The list of predictors presented to the classifier are 
the following:
\begin{itemize}
\item Modulation index, defined as:
\begin{equation}
\textrm M = \frac{
  \sqrt{\langle\textrm{I(}\nu\textrm{,t)}^2\rangle_{\nu,\textrm{t}} - 
    \langle\textrm{I(}\nu\textrm{,t)}\rangle_{\nu,\textrm{t}}^2}}
  {\langle\textrm{I(}\nu\textrm{,t)}\rangle_{\nu,\textrm{t}}},
\end{equation}
where $\textrm{I(}\nu\textrm{,t)}$ is the intensity in the event window\footnote{The 
event window is defined as the dedispersed frequency-time matrix, 
where the DM and width of the event window are chosen to optimally maximise S/N.} of the candidate. 
A time-averaged modulation index is also computed, described as the following:
\begin{equation}
\overline{\textrm M} = \frac{
  \sqrt{\langle\overline{\textrm I(\nu)}^2\rangle_{\nu} -
    \langle\overline{\textrm{I}(\nu)}\rangle_{\nu}^2}}
  {\langle\overline{\textrm I(\nu)}\rangle_{\nu}},
\end{equation}
where $\overline{\textrm I(\nu)}=\langle \textrm{I}(\nu,\textrm{t}) \rangle_{\textrm{t}}$ 
is the time-averaged spectrum of the FRB candidate.
\item The width of the candidate in data samples.
\item Fraction of power in each of the 3 RFI-dominated 5 MHz bands, centred at 842.5, 837.5 and 832.5 MHz:
\begin{equation}
    Fp_i = \frac{\sum_{\nu_s^i}^{\nu_e^i}\sum_{t} \textrm I(\nu,t)}{\sum_{\nu}\sum_{t} \textrm I(\nu,t)},
\end{equation}
where $\nu_s^i$ and $\nu_e^i$ are the start and end frequencies of each of the RFI bands.
\item The statistics and the $p$-values of the Kolmogorov-Smirnov and Shapiro-Wilk tests, 
 comparing the time-averaged spectrum to a normal distribution.
\item The mean ($\upmu$) and standard deviation ($\sigma$) of the event window.
\item The mean and standard deviation of windows with the same widths before and after the event window.
\item The ratio of number of pixels with intensity values greater than the mean, the mean plus one, and plus two times  
the standard deviation of the event window, to the total number of pixels in the event window, i.e.,
\begin{equation}
    f_i = \frac{N(\textrm I (\nu,t) > \upmu + i\sigma)}{N(\textrm I(\nu,t))},
\end{equation}
where $i={0,1,2}$ and $N(\textrm I(\nu,t))$ is the total number of pixels in a given event window.
\end{itemize}

\subsection{Validation} \label{subsec:validation}
When the model was first deployed on the live system of UTMOST, 
the pulsar catalog used for candidate cross-checking only consisted of pulsars 
that were already present in the training set. 
Single pulses from pulsars not listed in that catalog are treated 
as candidates and are presented to the classifier for evaluation. 
Observers would then receive email notifications of `new' detected pulsars, and, upon a user's validation, 
the catalog is appended with the pulsar names. 
More than 130 pulsar have been blindly `discovered' by the pipeline.
Over 250,000 pulses (excluding those from the bright pulsars Vela and J1644--4559) have been detected during
the survey.

In order to better understand the detection completeness of our system, 
we have developed a live injection system of simulated FRBs. 
A set of mock FRBs with a known set of S/N, DM, width, and scattering 
properties are held in a database on disk. 
The current mock injection algorithm operates in total power (detected data) space, and injections are performed directly on live data streams of individual fan-beams. 
In Fig.\,\ref{fig:injections}, we show the distribution of S/N, DM and width for the 
$\sim 2000$ injected FRBs (blue) and FRBs missed by our pipelines (red). 
The fake FRB parameter space was sampled uniformly in the S/N range of [9,50], DM of 
[50,5000] \dm\ and width [0,16] ms. 
Due to computational constraints, 
we did not sample the region with width$<$16 ms as thoroughly as width$>$16 ms. However, 
we do expect that the efficiency of our pipelines to decrease with increasing pulse widths.
In general, we do not see any obvious trends in the missing fraction of fake FRBs, and 
work is in progress to reduce the false negative rate of our pipelines.
Ninety per cent of the $\sim 2000$ injected FRBs were blindly recovered, establishing our 
confidence in the overall detection and classification pipelines. 
Plans are currently set to extend the algorithm to be able to inject FRBs in the complex-sampled data output of individual UTMOST modules. The main advantages 
are that mock FRBs injected at the voltage level have to pass through more of UTMOST's processing pipeline, such as the delay engine, RFI mitigation subroutine, and 
the beamformer.




\section{FRB discoveries}\label{sec:frb_discoveries}

\begin{figure}
  \centering
  \includegraphics[angle=-90, trim = {25mm 20mm 15mm 20mm}, clip, width=0.92\columnwidth]{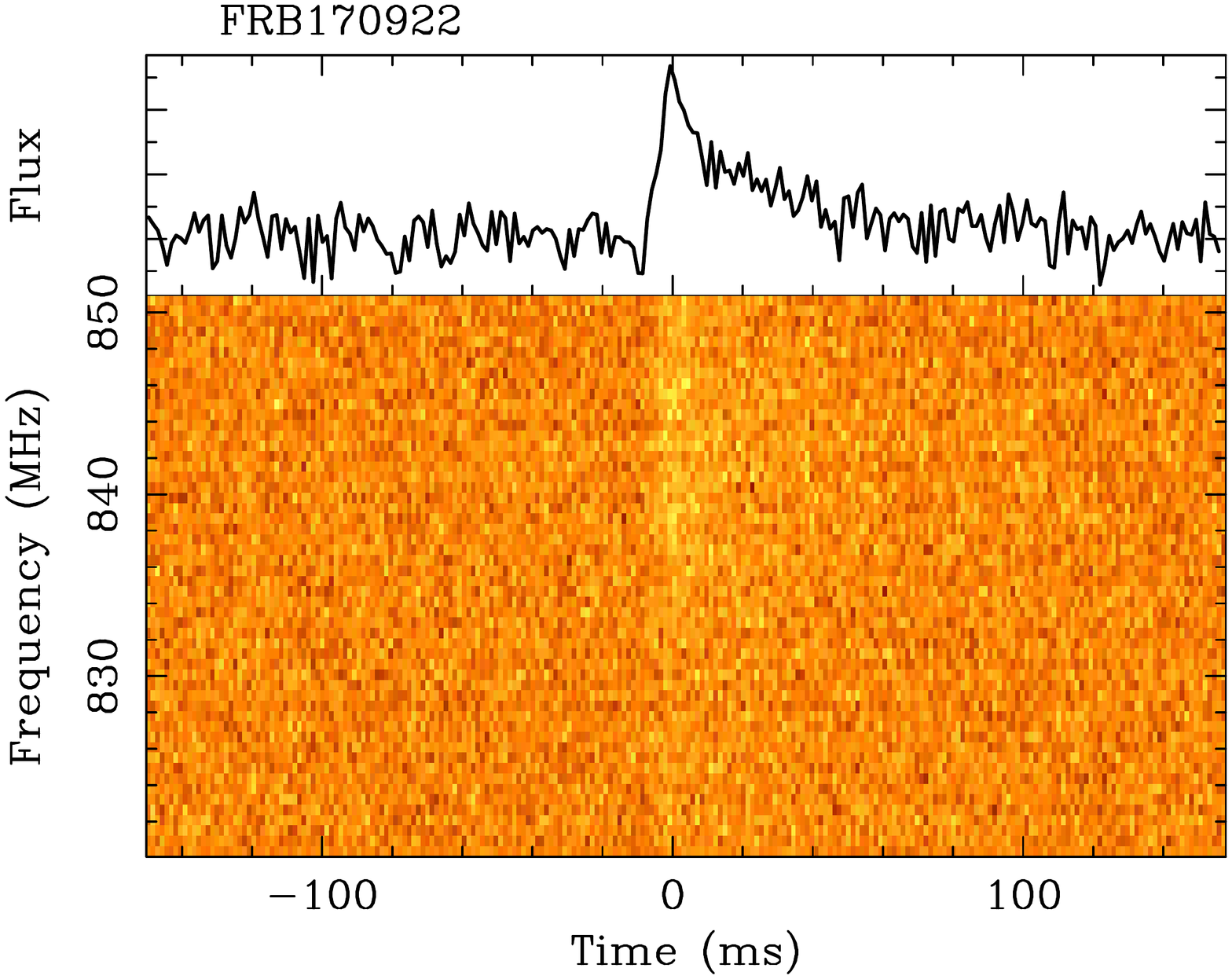}
  \includegraphics[angle=-90, trim = {25mm 20mm 15mm 20mm}, clip, width=0.92\columnwidth]{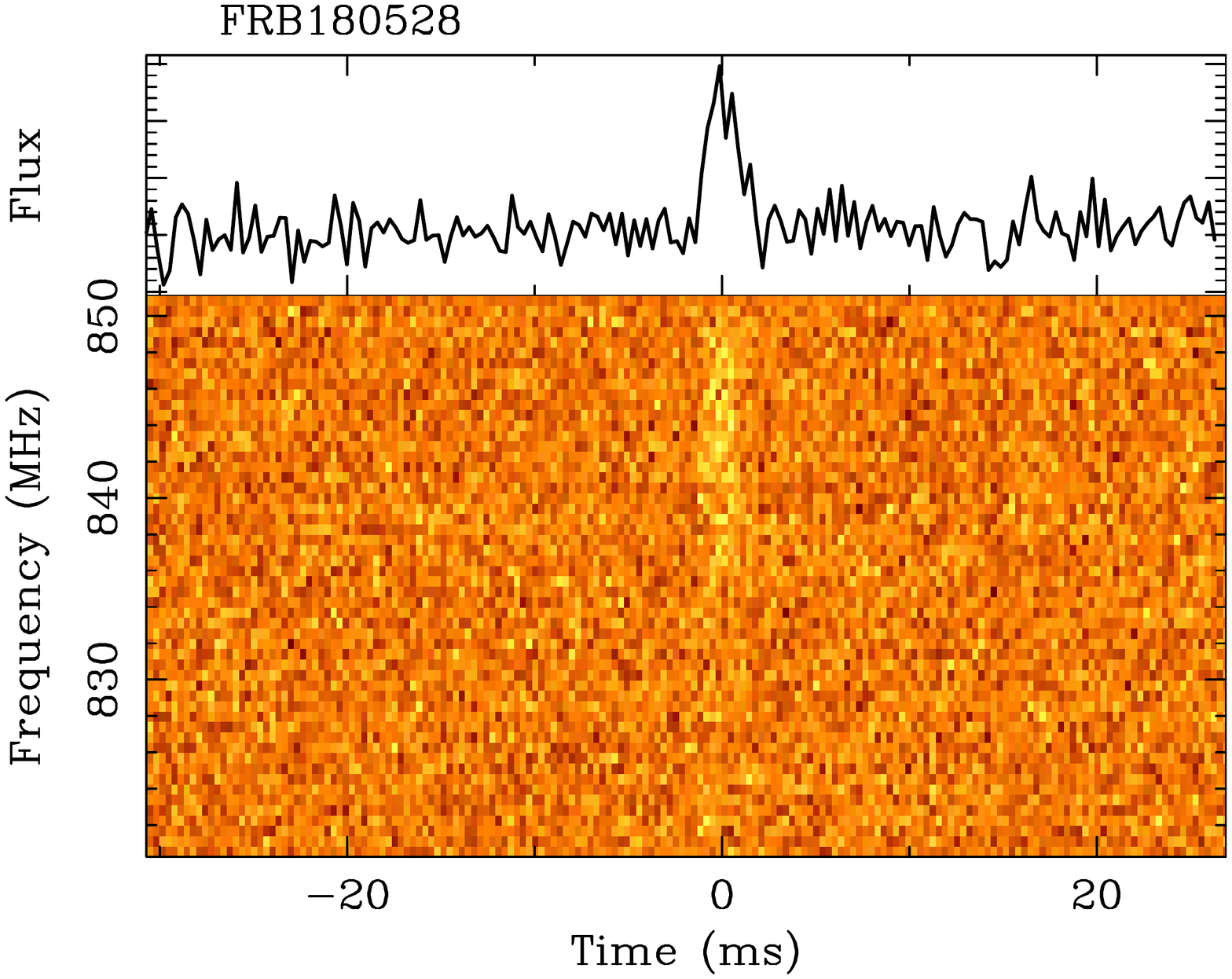}
  \includegraphics[angle=-90, trim = {25mm 20mm 15mm 20mm}, clip, width=0.92\columnwidth]{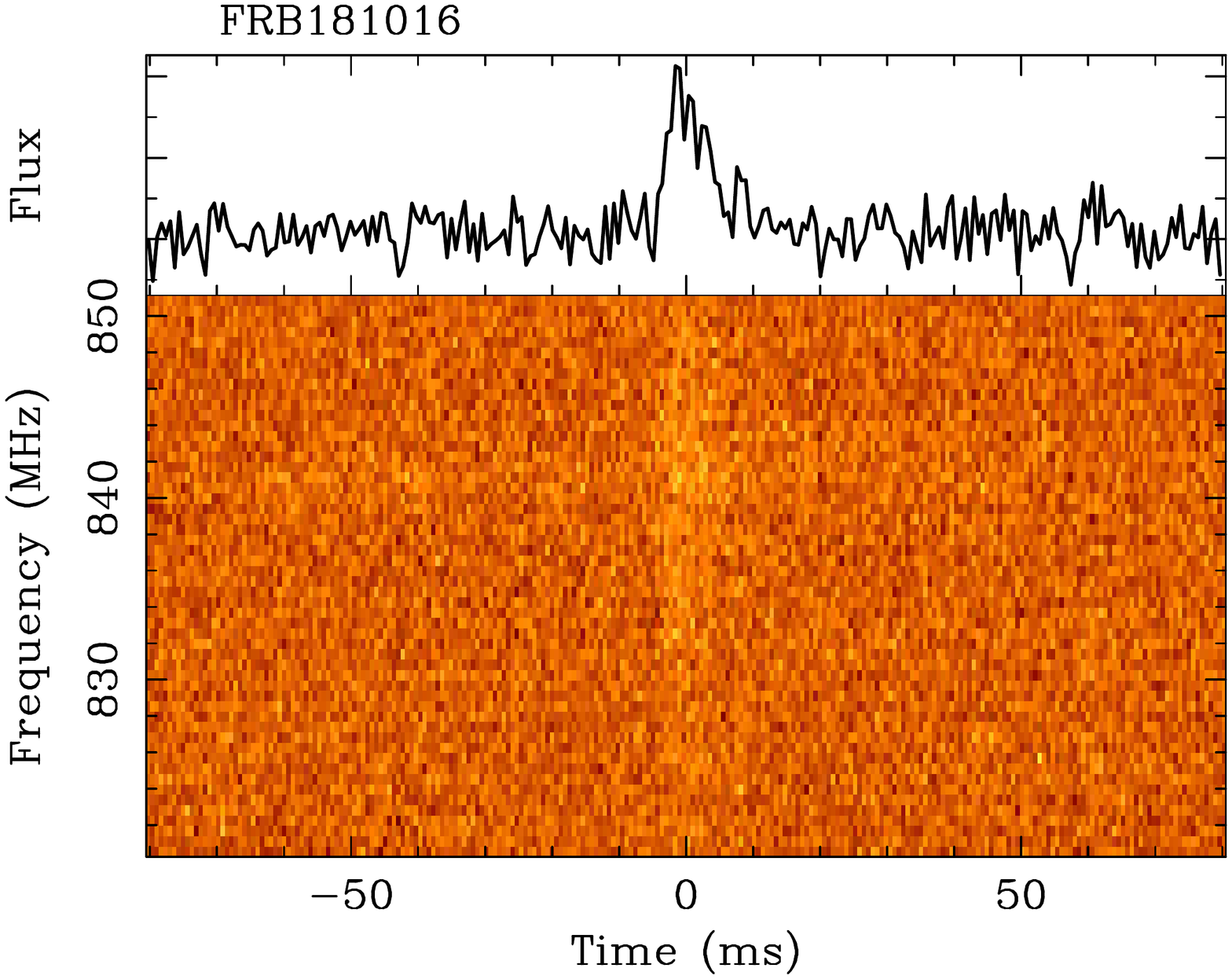}
  \includegraphics[angle=-90, trim = {25mm 20mm 15mm 20mm}, clip, width=0.92\columnwidth]{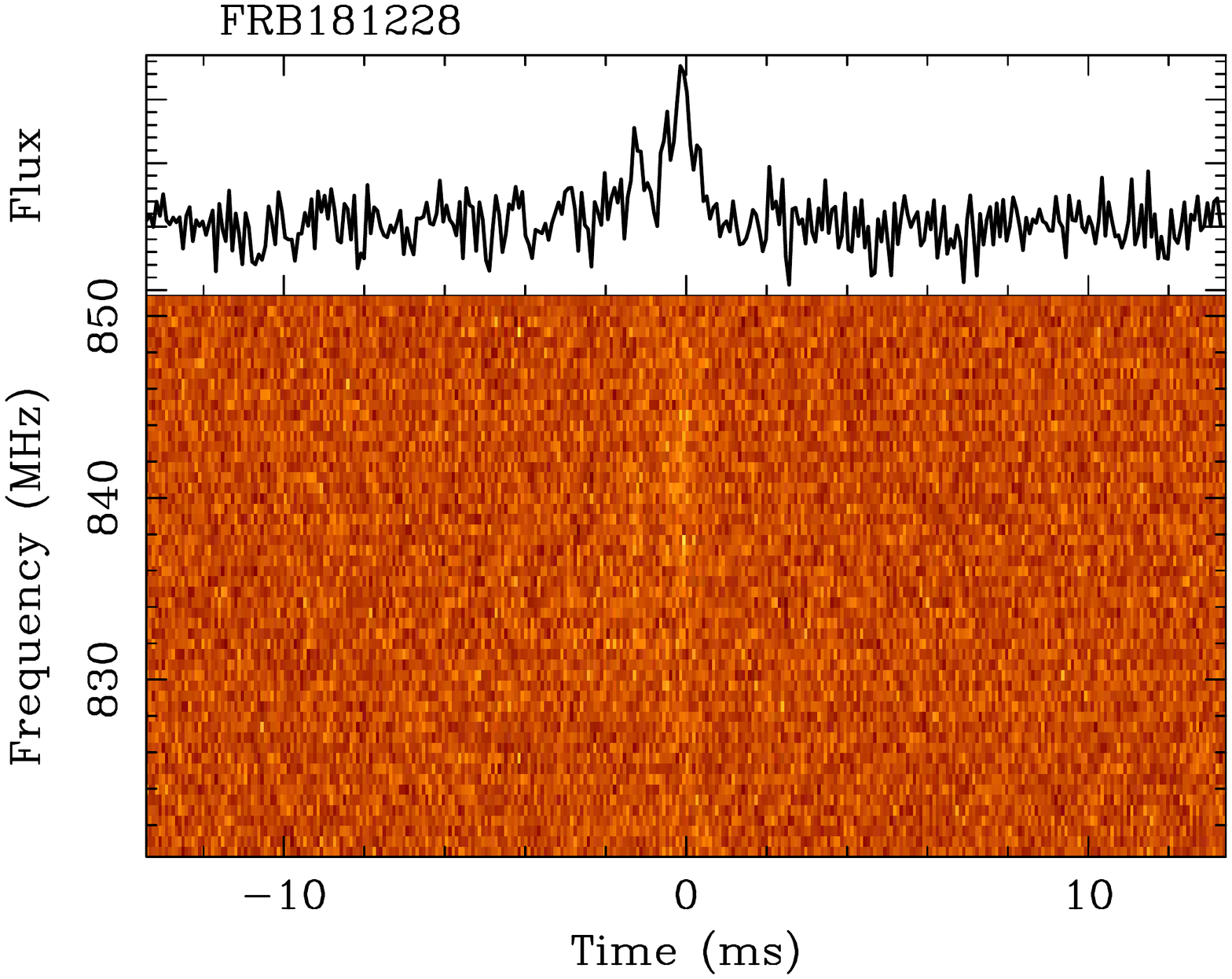}
  \caption{Dynamic spectra of FRB170922, FRB180528, FRB181016 and FRB181228. FRB170922 shows the largest scattering tail 
  measured for a fast radio burst with $\tau_{d}=29.1^{+2.8}_{-2.6}$\,ms. We note that UTMOST's resonant cavity is more
  sensitive in the range 835-850 than 820-835 MHz.}
  \label{figure:All_FRBs_spectra}
\end{figure}

\begin{figure*}
  \centering
  \includegraphics[trim = {20mm 15mm 20mm 15mm}, clip, width=0.8\textwidth]{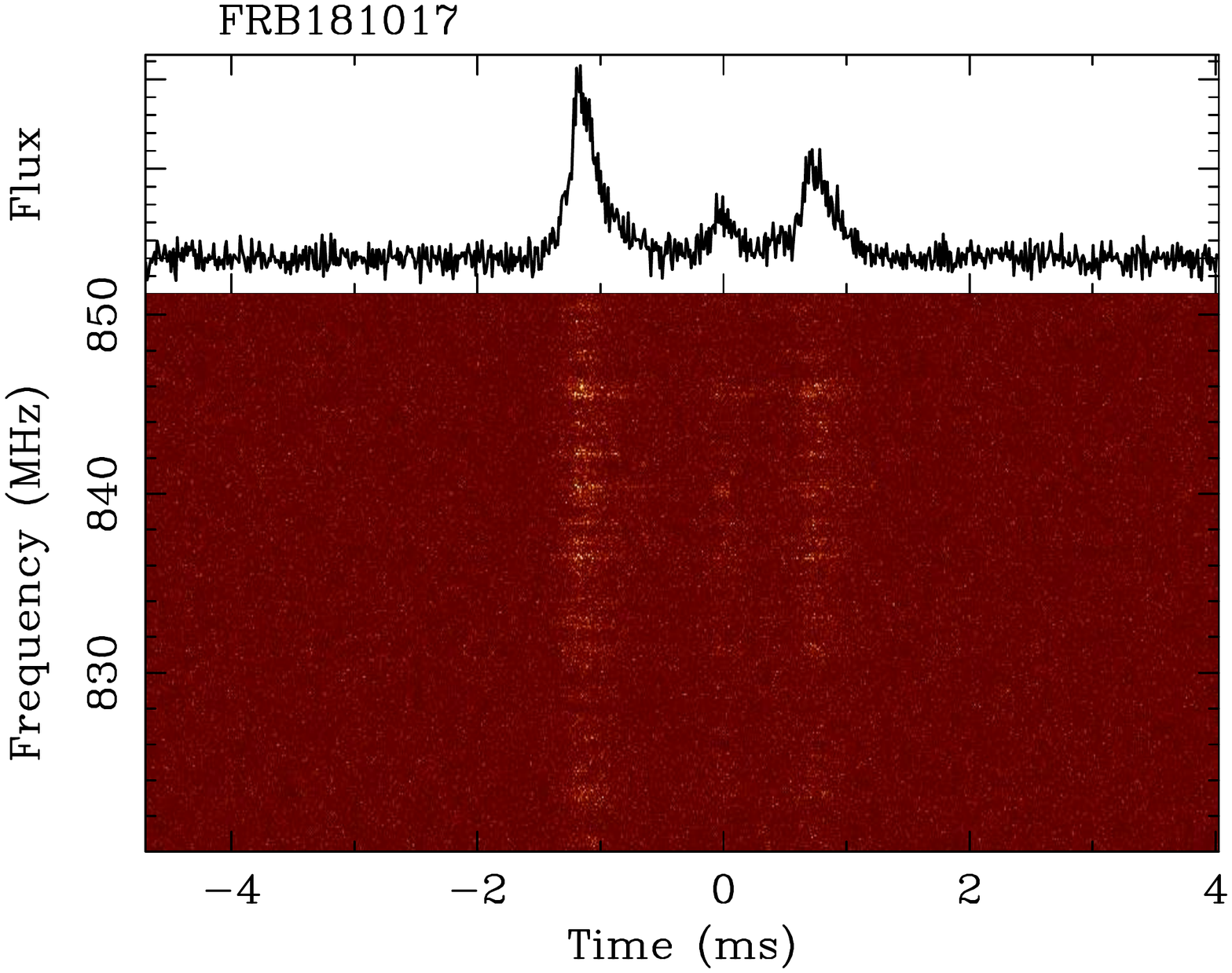}
  \caption{FRB181017: the triple-peaked FRB. The waterfall plot for the FRB is shown for frequency as a function of time. Voltage capture of the event yields much higher time resolution (10 $\upmu$sec) than we obtain from the off-line pipeline (655.36 $\upmu$sec). The frequency resolution is 97.66 kHz. The event shows a remarkable three peaked structure, with a spectrum which is quite similar across the peaks, similarly to what is seen in FRB170827. The three peaks have consistent scattering timescales and pulse widths. This scattering timescale would be associated with frequency structures at the kHz scale, far below the instrumental resolution. The striations in frequency are on scales of a few 100 kHz, and could be associated with the ISM (the NE2001 model predicts scintillation bandwidths at the position of the FRB of $\approx$ 2 MHz), although we cannot rule out they arise at the source or propagating through the host galaxy ISM and/or the IGM.}
  \label{figure:FRB181017}
\end{figure*}

Over 344 days of on-sky observations, the survey yielded six FRBs that passed our automatic and visual verification tests (Table\,\ref{table:FRBs}). 
One of these, FRB170827 has already been reported by \cite{Farah2018}. Here, we report  
the discovery of FRB170922, FRB180525, FRB181016, FRB181017 and FRB181228. All but one of these (FRB170922) were discovered in real 
time, where a voltage capture was triggered, allowing for improved localisation in the EW 
direction and coherent dedispersion (see \citealt{Farah2018}). 
As part of our policy to publicise confirmed events, Astronomer's Telegrams were issued for all the above FRBs
\citep{Farah_FRB170922_atel, Farah_FRB180528_atel, Farah_FRB18101_atel, Farah_FRB181228_atel}. 
The dynamic spectra 
of the FRBs, and their frequency-averaged pulse profile are displayed in Fig.\,\ref{figure:All_FRBs_spectra} and Fig.\,\ref{figure:FRB181017}.

The localisation arc of the FRBs can be described as a second-order 
polynomial of the form:
\begin{equation}
    \label{eq:localisation_arc}
    \textrm{RA} = \textrm{RA}_0 +a (\textrm{Dec} - 
    \textrm{Dec}_0) + b (\textrm{Dec} - \textrm{Dec}_0)^2,
\end{equation}
where RA$_0$ and Dec$_0$ are the coordinates of the most probable location.
We list the times of arrival, coordinates and the corresponding localisation arc parameters, 
and properties of our FRB sample in Table\,\ref{table:FRBs}. The reported detection S/N represents the signal-to-noise 
ratio evaluated by the discovery algorithm, a value which is particularly valuable for source-count studies (see e.g. 
\citealt{James2018_source_count}). 
To compute flux densities, we use the radiometer equation:
\begin{equation}
\label{eq:radiometer}
\textrm{S}_{\textrm{peak}} = \eta\times \textrm{S/N}\times\frac{T_{\textrm{sys}}}{G\sqrt{\textrm{BW} \times W_{\textrm{eq}}}},
\end{equation}
where $\eta$ is the beam attenuation correction factor in the EW direction, $T_{\textrm{sys}}=330$K is the system temperature, 
and $G$ is the gain of the instrument, determined using the latest phase calibrator prior to each FRB detection, typically $\sim1.7$
K/Jy. BW $=31.25$ MHz is the bandwidth of the Molonglo radio telescope, and $W_{\textrm{eq}}$ is the  equivalent width of 
the bursts. The equivalent width of an FRB represents the width of a top hat with height and area equal to 
the amplitude and area of the burst pulse profile. 
Due to the unconstrained position of the bursts in the NS direction, the measured flux densities represent 
lower limits of the values assuming the bursts were observed close to the beam centre.

We follow \cite{Zhang2018} to compute the maximum DM-inferred redshift of FRBs, assuming that the contribution of the host galaxies of FRBs to their measured DM is DM$_\textrm{host}$ = 50 \dm.
We follow \cite{Hogg1999} to estimate the in-band isotropic energy of FRBs:
\begin{equation} 
E = \frac{4\pi  D_{\textrm{L}}^2 }{(1+z)^{1+\alpha}}F_{\nu_{c}}\textrm{BW},
\end{equation}
where $F_{\nu_{c}}$ is the fluence of the FRB, BW is the bandwidth of the observing instrument, $D_{\textrm{L}}$ is 
the luminosity distance, and $\alpha$ is the spectral index ($F\propto\nu^{\alpha}$). We adopt the following cosmology \citep{Planck2015}: H$_0=67.74$\,km\,s$^{-1}$\,Mpc$^{-1}$ as the Hubble parameter, $\Omega_b = 0.0486$, $\Omega_{\textrm{m}}=0.3089$ and $\Omega_{\Lambda}=0.6911$ as the baryonic matter, total matter and dark energy density parameters, respectively, and we make use of the cosmology calculator CosmoCalc \citep{CosmoCalc}.

\begin{figure}
    \centering
    \includegraphics[width=\columnwidth]{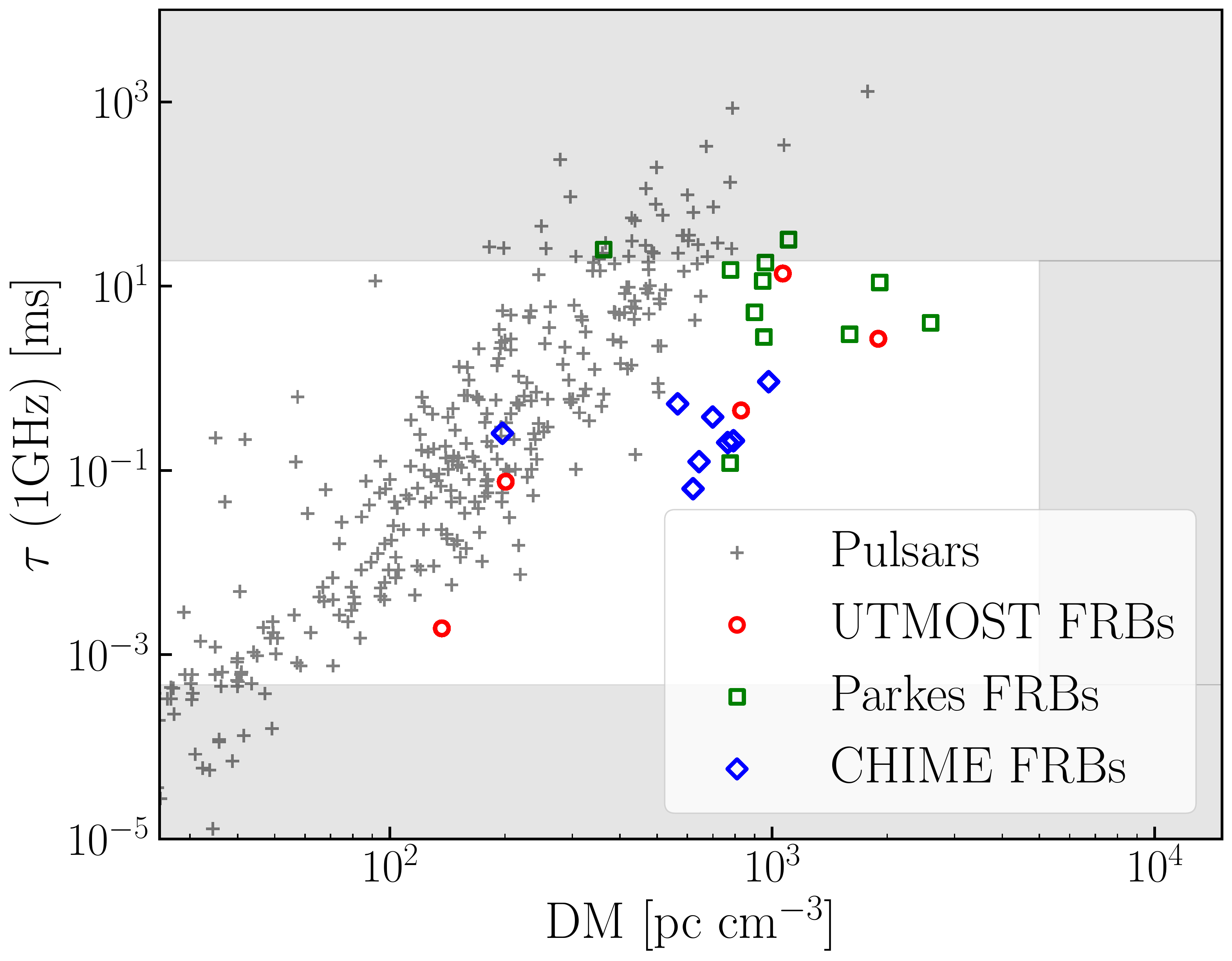}
    \caption{Broadening timescale as a function of the extragalactic DM for FRBs, and versus DM for Galactic pulsars. Only those FRBs are shown for which the scattering time can be measured; upper limits are not shown. The general trend is for FRBs to show less scattering than pulsars at the same DM. Note that the grey regions indicate approximately where we would expect to be strongly biased against finding FRBs: (1) because the DM limit is 5000 \dm on the UTMOST survey, with similar lower limits pertaining at the other two surveys, and (2) because of the time resolution limits (10 $\upmu$sec and $\approx 40$ ms for UTMOST). }
    \label{fig:tau_dm}
\end{figure}

A radio signal traversing turbulent media undergoes multi-path propagation, resulting in delayed times of arrival due to the 
additional light travel distance. This effect is evident as a trailing exponential tail on a dedispersed pulse profile. Pulse broadening is modelled as a Gaussian convolved with a one sided exponential 
of the form:
\begin{equation}
\label{eq:scattering_model}
\mathscr{M} = \textrm{A}\times\textrm{exp}\Big[\frac{-(t-t_0)^2}{2\sigma^2}\Big] \ast \Big\{\textrm{exp}\Big[-\frac{t-t_0}{\tau_{\textrm{d}}}\Big]\textrm{U}(t_{0})\Big\},
\end{equation}
with:
\begin{equation}
  \textrm{U}(t) =
  \begin{cases}
    0 & t<t_0 \\
    1 & t\geq t_0,\\
  \end{cases}
\end{equation}
where $\ast$ denotes convolution. $\tau_{\textrm{d}}$ is the scattering timescale, and $\sigma$ is the 
Gaussian width. Parameter estimation was performed using the \texttt{BILBY} package \citep{bilby}, making use of the \texttt{pyMultiNest} sampler 
\citep{pymultinest}. We used a Gaussian likelihood function for our parameter estimation, along with uniform priors 
on all the fitted parameters. The scattering timescale measurements as a function of 
extragalactic DM of our latest FRBs are plotted in red in Fig.\,\ref{fig:tau_dm}.
A major current advantage of UTMOST is the capacity to capture voltages for FRBs, permitting scattering tails to be 
resolved and measured for narrower events than the bulk of FRBs to date at other facilities. Highly 
scattered low DM FRBs are detectable in principle in all FRB surveys plotted in Fig.\,\ref{fig:tau_dm} 
but, to-date, have not been. When voltage 
capture becomes routine at other facilities, narrow but high DM events can be expected.

We show the observed and fitted profiles in Fig.\,\ref{figure:All_FRBs_fit}, the posterior 
distributions of the Gaussian widths and the scattering timescales are shown in Fig.\,\ref{figure:All_posteriors}. 
We note that all the FRBs presented here are over-scattered with respect to the expectation from the Milky Way along their 
lines of sight, according to the NE2001 model \citep{NE2001}.

\subsection{FRB170922}
\label{subsec:FRB170922}
FRB170922 has a measured DM  of 1111 \dm\,and shows a relatively large scattering tail, as can be seen in Fig.\,\ref{figure:All_FRBs_spectra}. We fit the 
profile using the above 
method and measure a scattering timescale of $29.1^{+2.8}_{-2.6}$ ms, one of the largest for an FRB.
FRB170922 was successfully discovered by UTMOST's live detection algorithm during a period of downtime, in which the system 
was recovering from a previous (false) trigger, which had taken place $\sim20$ seconds prior. 
The width of the 
FRB pulse is much larger than the inter-channel smearing time due to DM, and hence coherent dedispersion would have yielded no significant 
enhancement in S/N.

\subsection{FRB180528}
\label{subsec:FRB180528}
The coherently dedispersed pulse profile of FRB180528 at its DM of 899 \dm\,shows hints of temporal broadening at high time resolution. Fitting 
the profile with the model defined in Eq.\,\ref{eq:scattering_model}, we find that the scattering timescale at 835 MHz 
is $\tau_{d}=0.95^{+0.33}_{-0.35}$ ms, a value consistent with 0 at the 3-sigma level. 
This is evident in Fig.\,\ref{figure:All_posteriors} as the posterior distribution of $\tau_{d}$ is unbounded at 
the lower edge of the prior range ($\tau_{d}=0$).

\subsection{FRB181016}
\label{subsec:FRB181016}
FRB181016 represents the highest DM FRB that UTMOST has discovered to date, with a DM of 1984~\dm. 
The burst detection caused us to increase the DM threshold limit 
of the live pipeline from 2000 to 5000~\dm. Given the observed fluence 
and the relatively high DM, FRB181016 is inferred to be one of the most luminous FRBs, 
with an average inferred isotropic luminosity of $L\sim10^{44}$ erg/s. We measure a scattering 
timescale of 5.7 $\pm 0.8$ ms.

\subsection{FRB181017}
\label{subsec:FRB181017}
The dynamic spectrum of FRB181017 (Fig.\,\ref{figure:FRB181017}) reveals rich spectral and temporal structure. 
Unresolved in the detection filterbank due to the low time resultion, the high time resolution timeseries of FRB181017 
shows three burst peaks, separated in time by $\sim1$ ms. We note that the temporal separation of the leading and intermediate
peaks is larger than the separation between the intermediate and the trailing ones; thus, the episodic nature of the bursts 
cannot be explained by an underlying periodicity.

As the three peaks show hints of scattering, we fit the pulse profile by a model consisting of a summation of three Gaussian 
distribution functions with variable widths, convolved with the same exponential scattering timescale. 
We find that the (Gaussian) widths of the 
peaks are comparable, with a mean = $80\,\upmu$s, and the measured scattering 
timescale is $\tau_{\textrm{d}} = 160\,\upmu$s. We also fit the profile with a 
variable $\tau$ for each peak and find that the scattering is consistent between them.
We measure the decorrelation bandwidth by fitting the constructed spectral auto-correlation function with a Gaussian 
function as described in \cite{Farah2018}. We find that the decorrelation bandwidth is $\nu_d=0.36$\,MHz.


Given the resemblance in the temporal structure of the three features of the burst, we explore the 
hypothesis that the lagging peaks are copies of the leading one (e.g. \cite{munoz2016_lensing,Cordes2017}) by searching 
for correlation in voltages between them. 
From the saved raw voltages, we first create a complex-sampled filterbank at the native time and frequency 
resolution of the instrument by placing a tied array beam on the best known position of the FRB. The 
filterbank is then coherently dedispersed using 
a custom-built dispersion-removal software\footnote{\href{https://github.com/wfarah/pydada}{https://github.com/wfarah/pydada}}. 
A delayed signal traversing a different path might not encounter the
same electron density as the main pulse, and hence might be 
dispersed differently. A small difference in DM between the pulses might 
de-cohere the cross-correlation product. 
For example, if one pulse is dispersed 0.1 \dm\,more than the other, 
the expected delay in arrival times between them, at the bottom of the UTMOST band,  
is $\sim$ 40 $\upmu$s (or $\sim$ 4 time samples).
Hence, we perform a grid search over DM by coherently dedispersing one of the pulses 
$\pm 2$ \dm\ with respect to the other, in steps of $0.01$ 
\dm\ prior to cross-correlation.

For each frequency channel, we compute the cross-correlation 
of the dedispersed voltage stream $e(\nu,t,\textrm{dm})$ with a delayed copy 
of itself that has been trial dedispersed, $e(\nu, t+\delta t,\textrm{dm}+\delta \textrm{dm})$:
\begin{equation}
\textrm{V}(\nu, \delta t) = \langle e(\nu,t,\textrm{dm})e^*(\nu, t+\delta t,\textrm{dm}+\delta \textrm{dm})\rangle,
\label{eq:cross-corr}
\end{equation}
where $^*$ represents the complex conjugate operator, and angular brackets denote time averaging.
We select a windowing function that is approximately equal to the width of a single peak, and we search
in the range $-500 < \delta t < 500$ time samples.
For every sample delay $\delta t$, we calculate the \textit{degree of coherence},
\begin{equation}
\gamma(\delta t) = \frac{\widetilde{\textrm{V}}(t,\delta t)}{\langle e(\nu,t)e^*(\nu,t)\rangle}
                 = \frac{\widetilde{\textrm{V}}(t,\delta t)}{|e(\nu,t)|^2} ,
\end{equation}
where $\widetilde{\textrm{V}}(t,\delta t)$ is the lag spectrum computed by taking the inverse Fourier 
transform of $\textrm{V}(\nu,\delta t)$, and the denominator represents the amplitude of the auto-correlation 
function. 
In the limiting cases, the two temporal peaks of FRB181017 at any given $\delta t$ would be completely coherent (incoherent) if 
$|\gamma(\delta t)| = 1 (0)$. 
We found no evidence that the temporal 
features of FRB181017 are phase correlated
by placing a 5-$\sigma$ upper limit of 2.5$\%$ on the degree of coherence between the 
three FRB peaks.
We conclude that this 
triple-peaked structure is most likely intrinsic to the source emission.

\subsection{FRB181228}
\label{subsec:FRB181228}
\begin{figure*}
  \centering
  \includegraphics[width=\columnwidth, clip]{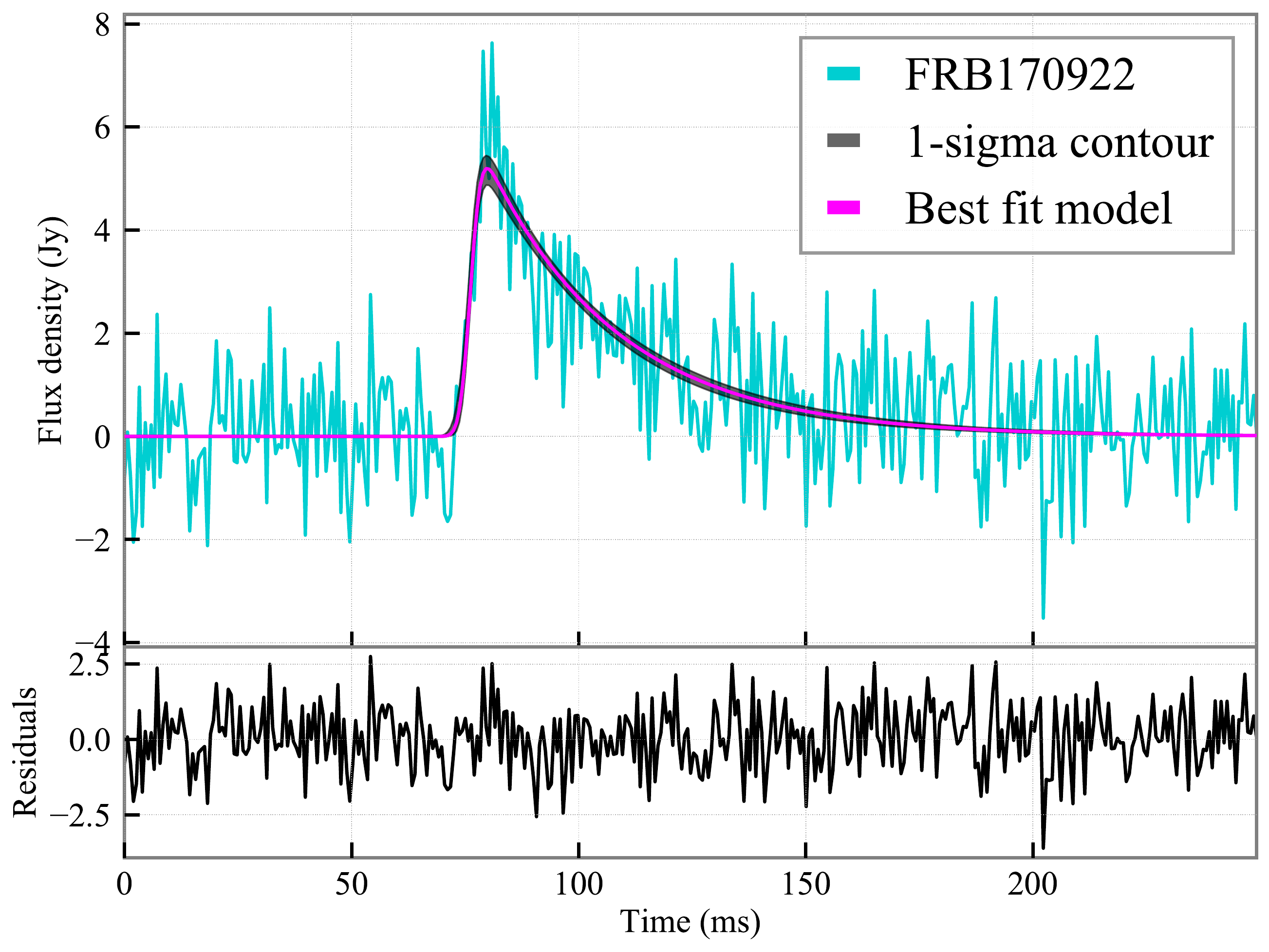}
  \includegraphics[width=\columnwidth, clip]{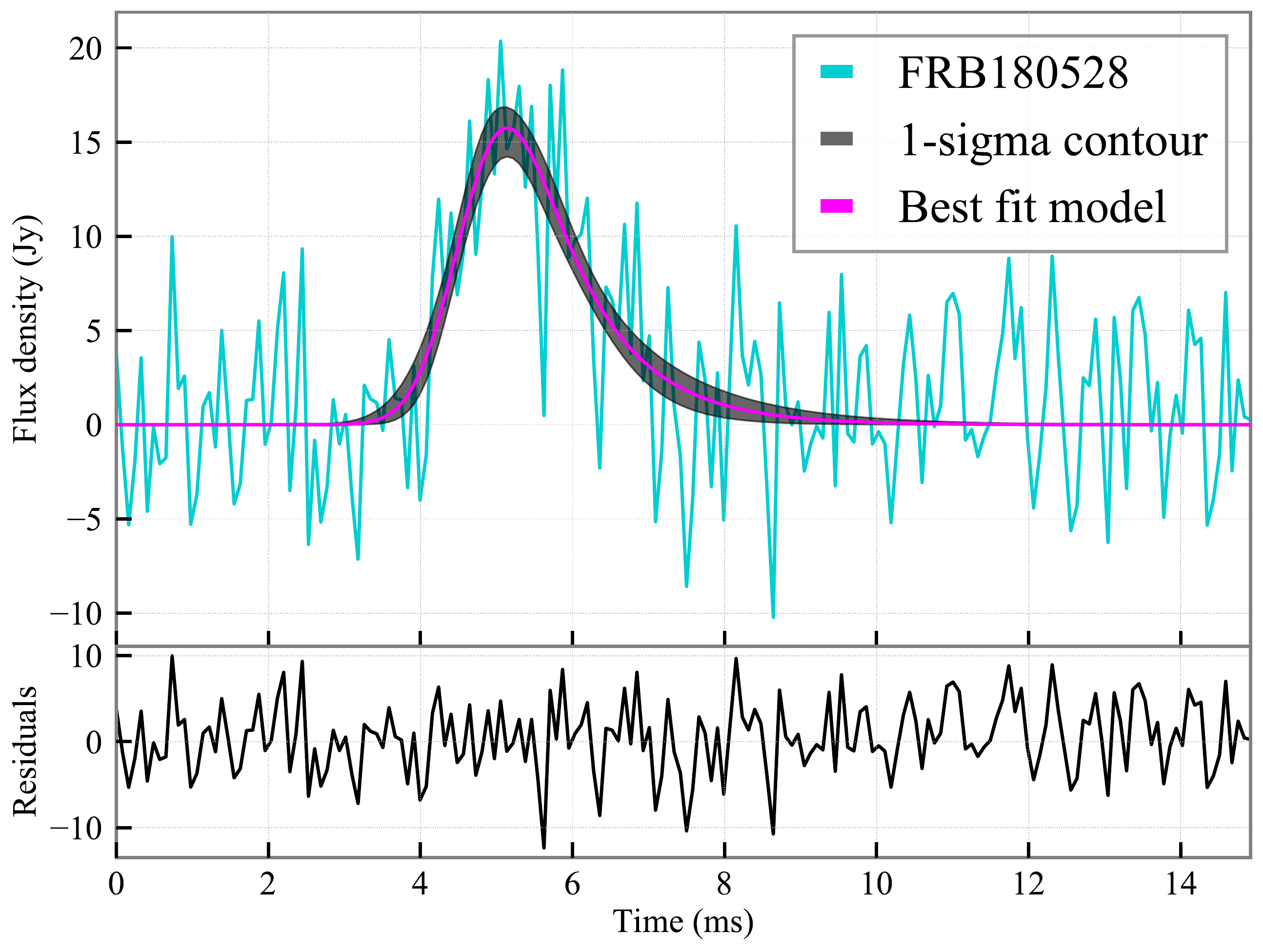}
  \includegraphics[width=\columnwidth, clip]{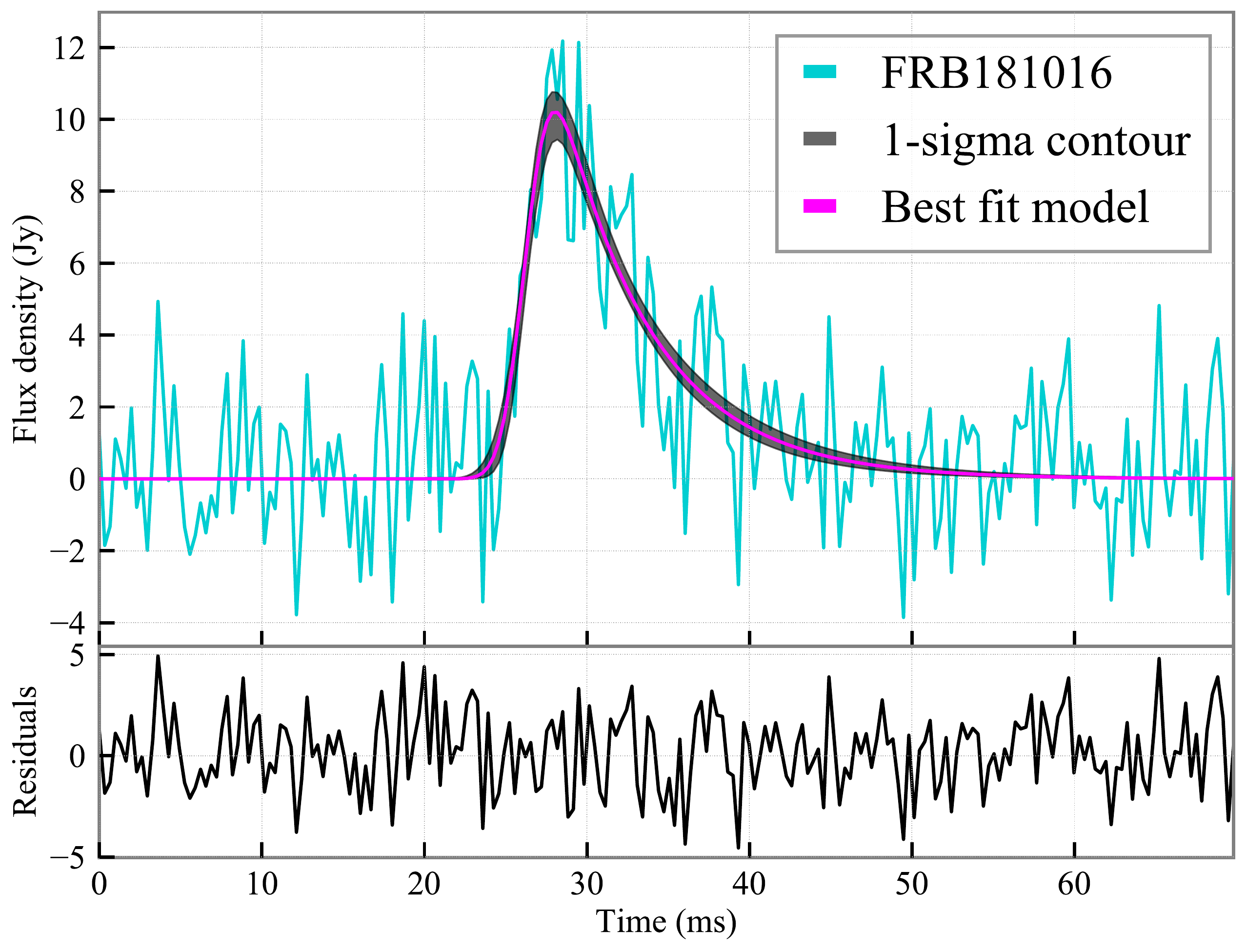}
  \includegraphics[width=\columnwidth, clip]{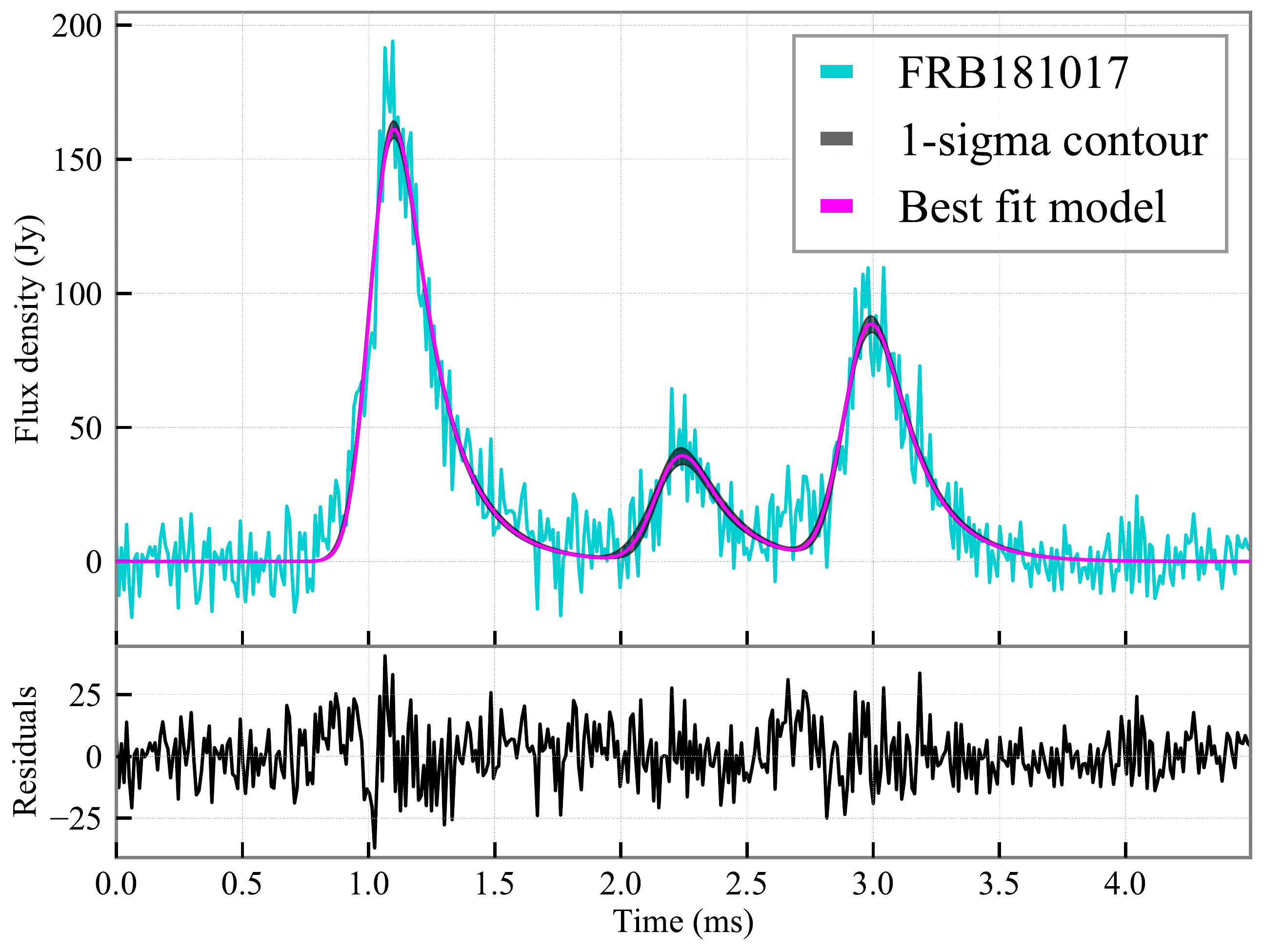}
  \includegraphics[width=\columnwidth, clip]{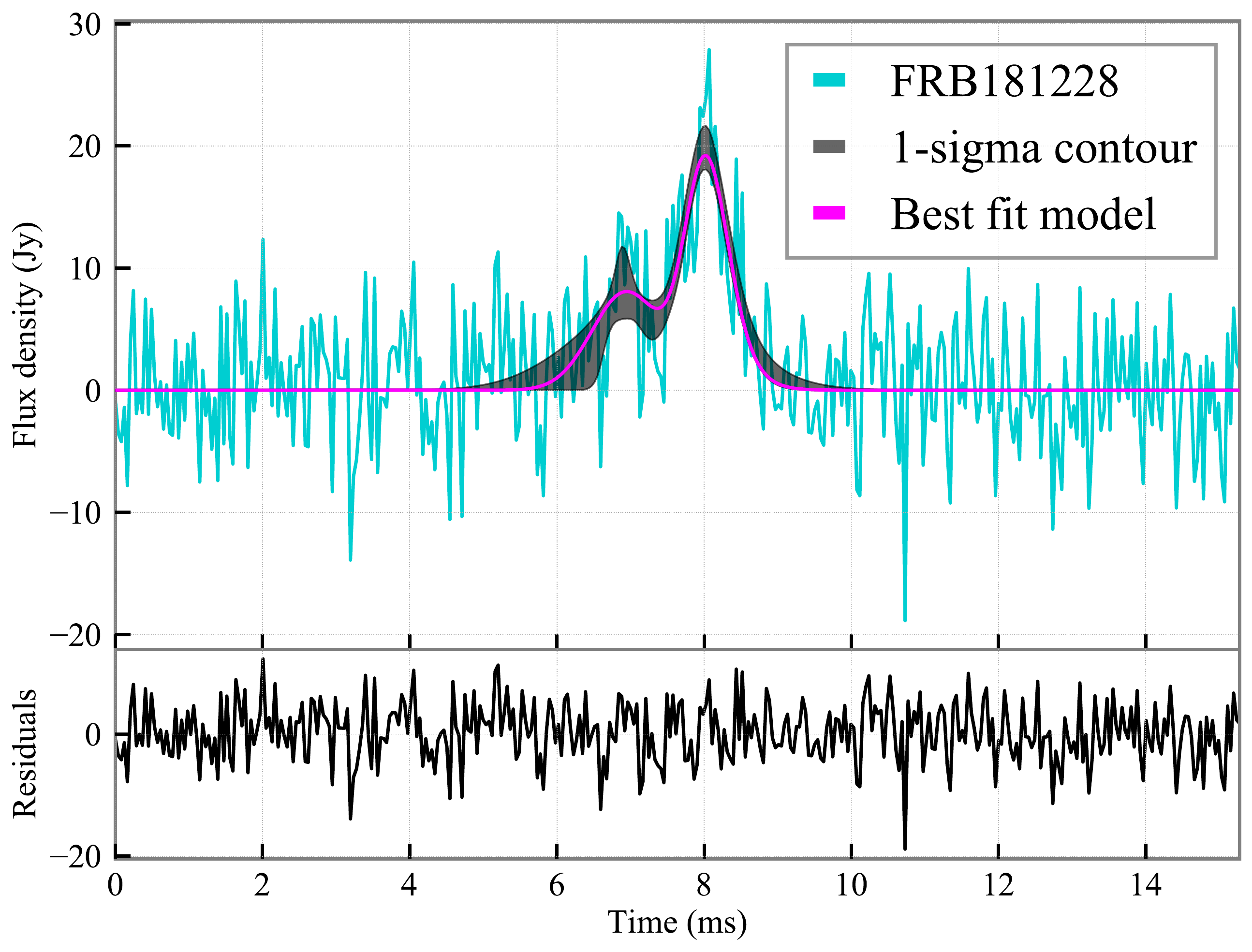}
  \caption{Frequency-averaged timeseries of the 5 FRBs presented in this paper. 
  The timeseries of FRB170922, FRB180528 and FRB181016 are fitted with the 
  model described in Eq.\,\ref{eq:scattering_model}, whereas FRB181017 and FRB181228 
  are fitted with a modified model (see \cref{subsec:FRB181017} and \cref{subsec:FRB181228}).}
  \label{figure:All_FRBs_fit}
\end{figure*}

\begin{figure}
\centering
  \subfloat[FRB170922.\label{fig:1a}]{\includegraphics[width=0.48\columnwidth]
  {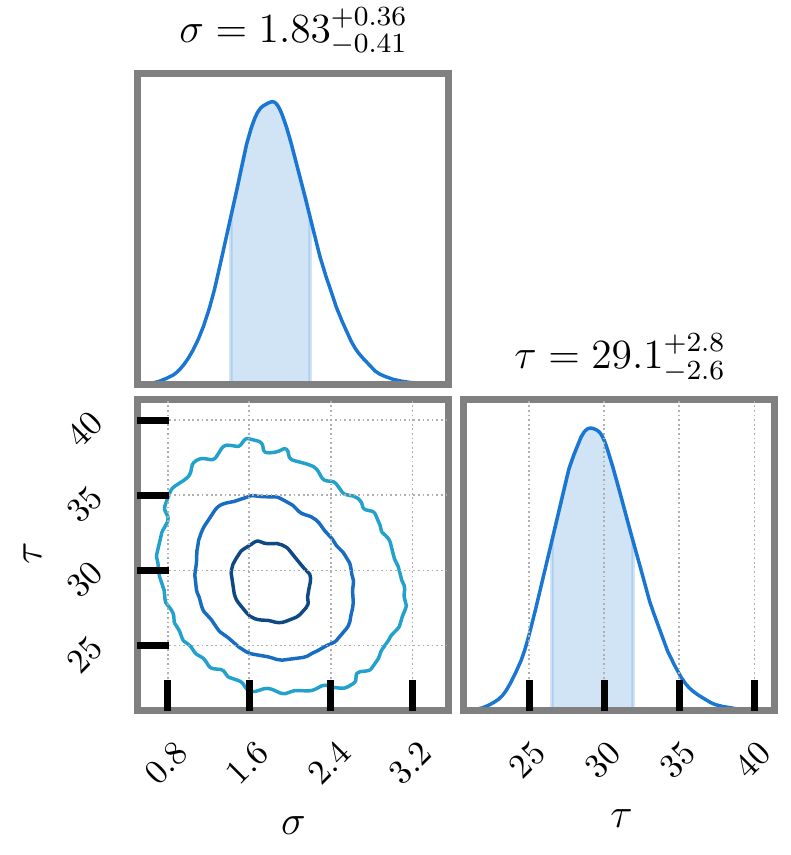}}
  \subfloat[FRB180528.]{\includegraphics[width=0.48\columnwidth]
  {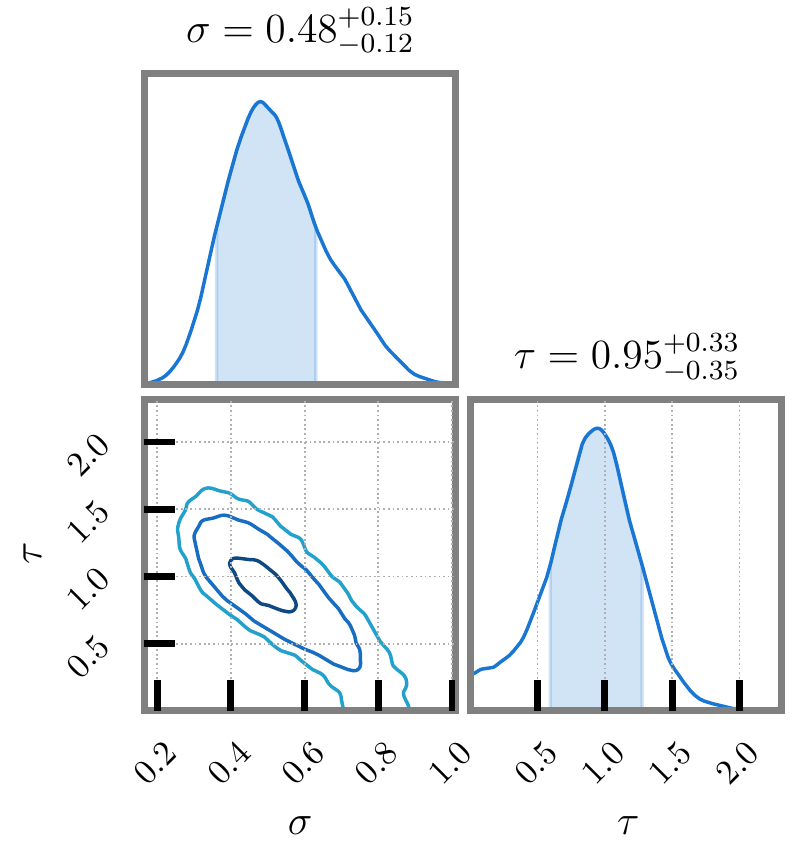}}\hfill
  \subfloat[FRB181016.]{\includegraphics[width=0.48\columnwidth]
  {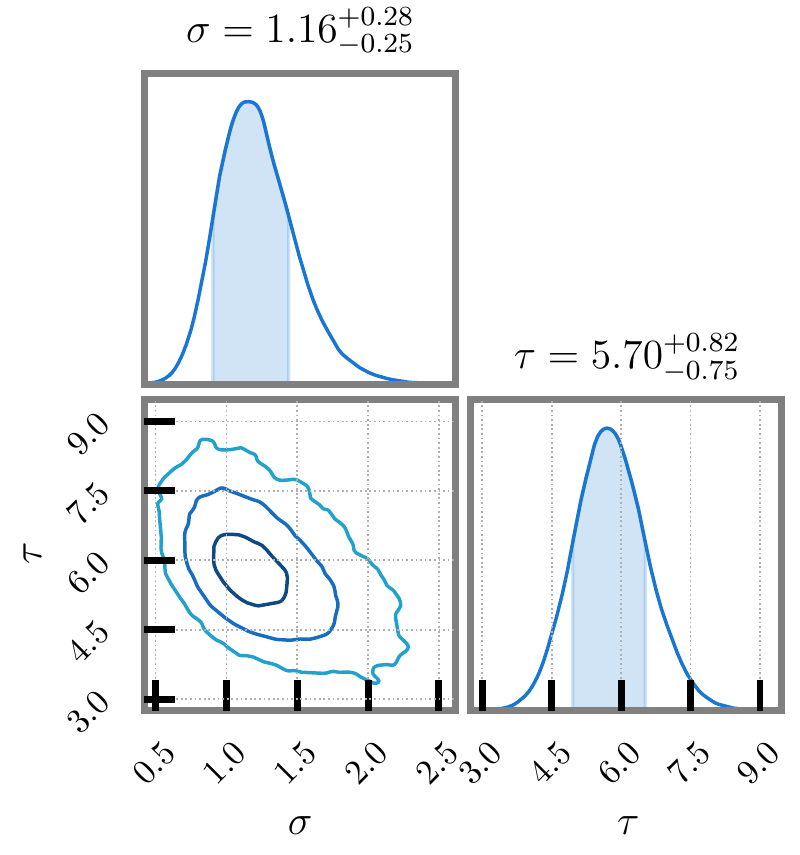}}
  \subfloat[FRB181017.]{\includegraphics[width=0.48\columnwidth]
  {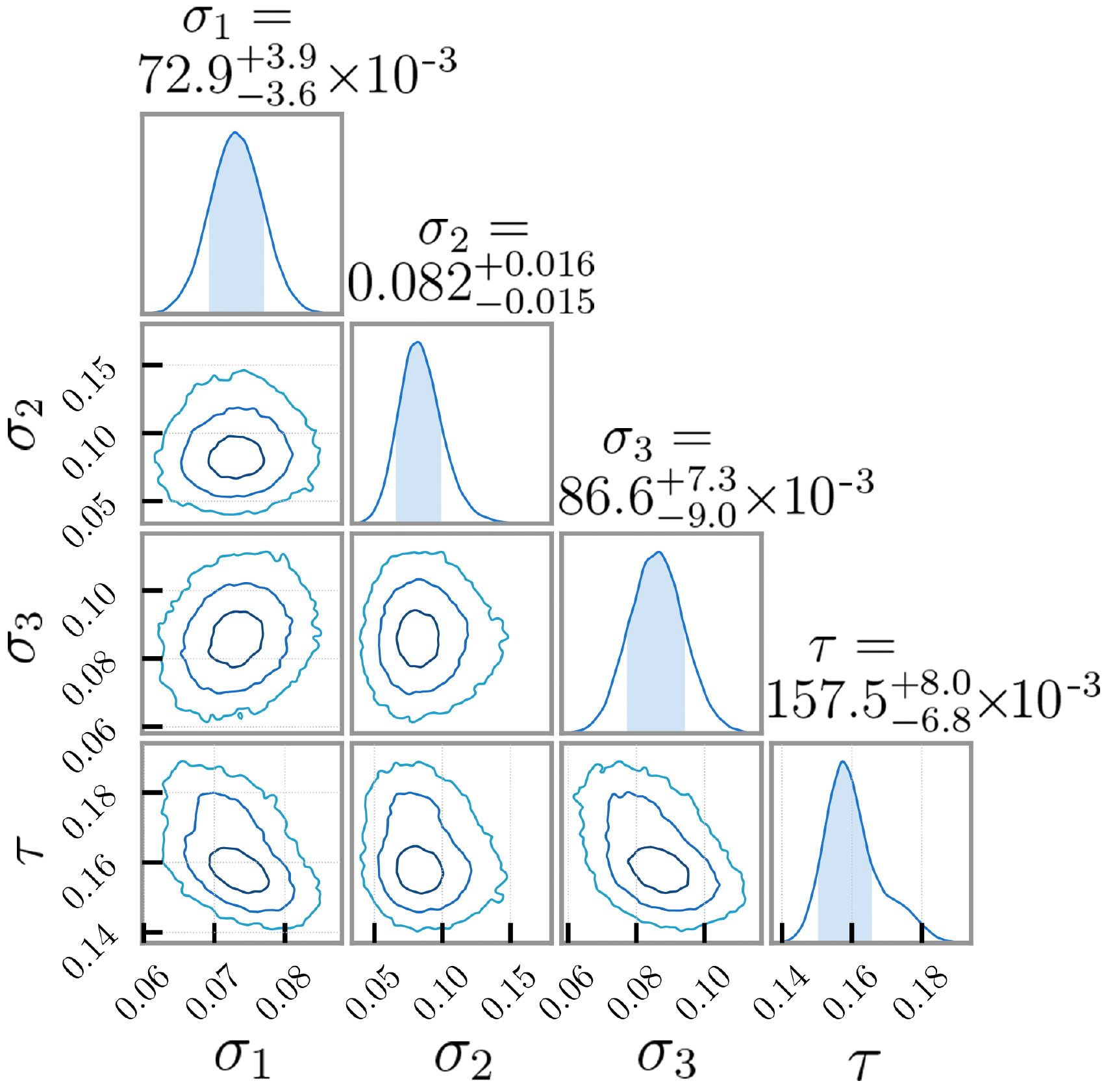}}\hfill
  \subfloat[FRB181228.]{\includegraphics[width=0.48\columnwidth]
  {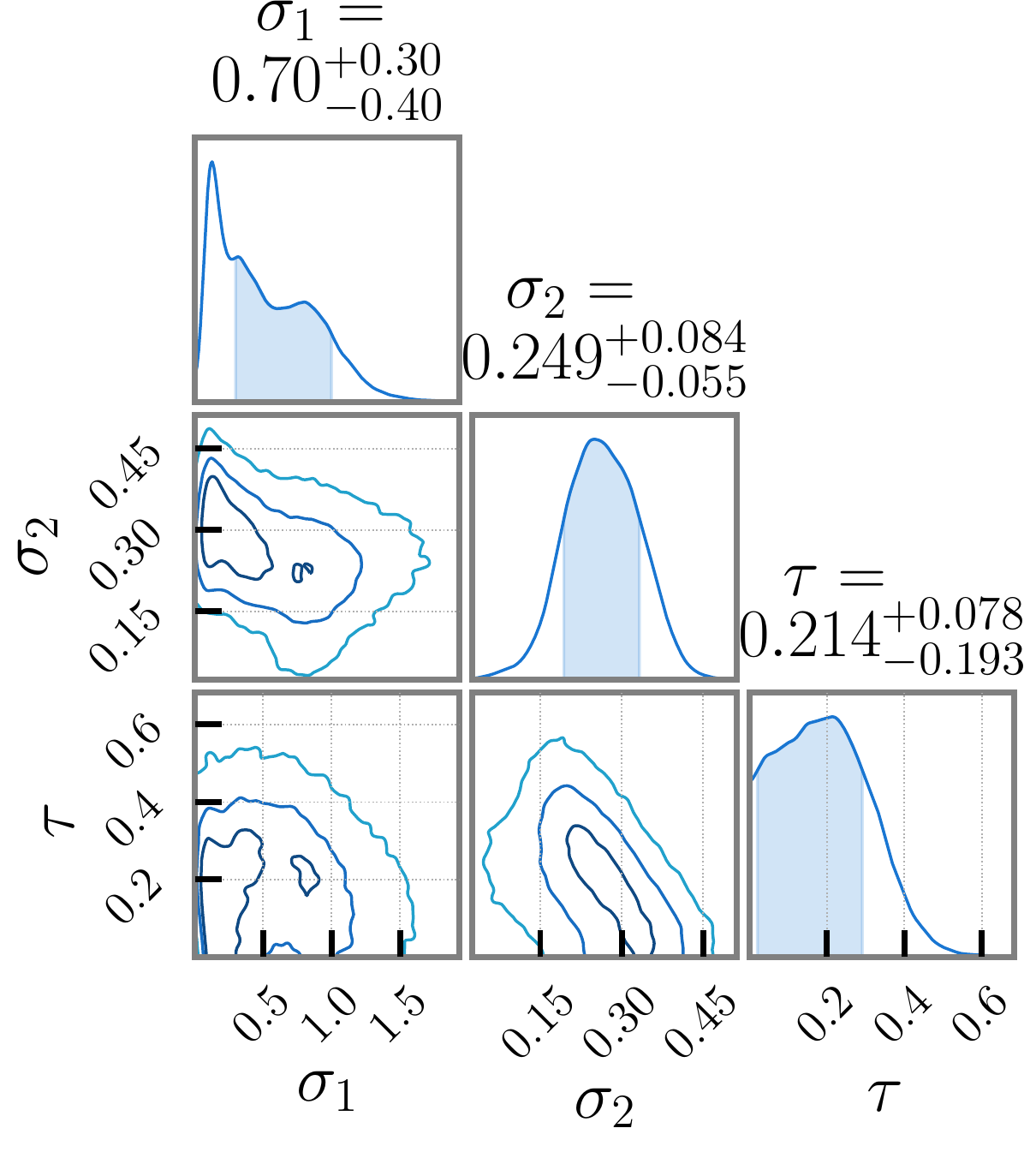}}
  \caption{The posterior distributions of the Gaussian width ($\sigma$) and scattering timescale ($\tau_{d}$) for the fitted 
    model (Eq.\,\ref{eq:scattering_model}) for the new FRBs reported. 
    All values shown are in ms.}
  \label{figure:All_posteriors}
\end{figure}

A hint of a precursor is visible in the dynamic spectrum and the dedispersed timeseries of FRB181228 as seen in Fig. \ref{figure:All_FRBs_spectra}. Similar to FRB181017, the pulse profile of FRB181228 was modelled using two Gaussians convolved with an exponential. 
The modelling of the pulse profile of this FRB proved challenging due to its low S/N evidenced by a large 1-sigma contour in the fit  (Fig.\,\ref{figure:All_FRBs_fit}) and its unbounded posteriors  (Fig.\,\ref{figure:All_posteriors}).
As the measured $\tau$ is consistent with being zero at the 2-sigma level ($\tau=0.21^{+0.08}_{-0.19}$ ms), we consider this measurement as an upper-limit.

%
         

\begin{table*}
  \centering
  \caption{Arrival times, coordinates and the properties of the FRBs 
  reported in this paper. The coordinates (RA, Dec) and (\textit{Gl, Gb}) 
  represent the centre of the localisation arc described in Eq.\,\ref{eq:localisation_arc}.}
  \label{table:FRBs}
  \begin{adjustbox}{max width=\textwidth} 
    \begin{tabular}{l c c c c c}
      \hline
      \ & \textbf{FRB179022} & \textbf{FRB180528}  & \textbf{FRB181016} & \textbf{FRB181017} & \textbf{FRB181228} \\ 
      \hline
      & & & \textbf{Arrival time and coordinates} & & \\
     \hline
         Event time at 850 MHz UTC & 2017-09-22 11:23:33.4 & 2018-05-28 04:24:00.9 & 2018-10-16 04:16:56.3 & 2018-10-17 10:24:37.4 & 2018-12-28-13:48:50.1\\

         RA, Dec (J2000) &  21:29:51.22, -07:59:40.48 & 06:38:49.80, -49:53:59.0  
         & 15:46:20.84, $-$25:24:32.6  & 22:05:54.82, $-$08:50:34.22 & 06:09:23.64, $-$45:58:02.4 \\
         \textit{Gl, Gb}  & 45.0683$^{\circ}$, $-38.7006^{\circ}$ &  $258.8723^{\circ}$, $-22.3530^{\circ}$ &  
         $345.5101^{\circ}$, $+22.6607^{\circ}$ &  50.0564$^{\circ}$, $+46.8816^{\circ}$ & 253.3519$^{\circ}$, $-26.1469^{\circ}$\\
         
         RA$_0$ (hours) &  21.497561  & 6.647167 &   15.772456 & 22.098561 & 6.156567 \\
         Dec$_0$ (degrees) & $-7.994578$ & $-49.899722$ & $-25.409056$ & $-8.842839$ & $-45.967333$ \\
         
         $a^{\ast}$ & $3.808988\times10^{-5}$ & 1.398369$\times 10^{-3}$ &   $9.641957\times10^{-4}$   & $1.162544\times10^{-4}$ & $1.057078\times10^{-3}$ \\
         $b^{\ast}$ & $1.056944\times10^{-5}$  & $-3.903834\times10^{-5}$ &   $-2.033868\times10^{-5}$  & $4.269530\times10^{-6}$ & -$2.784623\times10^{-5}$ \\
         Dec range & $[-12,-4]$ & $[-54, -46]$ & $[-30,-21]$ & $[-13,-5]$ & $[-50,-42]$\\ 
      \hline
      &  & & \textbf{Measured Properties}  & & \\
      \hline
      \textsc{heimdall} detection S/N & 20 & 11 & 17 & 66 & 12\\
      Dispersion measure, DM (\dm) & 1111$\pm 1$ & 899.3 $\pm 0.6$ & 1982.8 $\pm 2.8$ & 239.97 $\pm 0.03$ & 354.2 $\pm 0.9$\\
      Scattering time at 835~MHz (ms) & 29.1$^{+2.8}_{-2.6}$ & 1.0$^{+0.3}_{-0.4}$ & 5.7$^{+0.8}_{-0.8} $ & 158$^{+8}_{-7}\times10^{-3}$ & $<$0.2\\
      Gaussian width (ms) & 1.8$^{+0.4}_{-0.4}$ & 0.5$^{+0.2}_{-0.1}$ & 1.2$^{+0.3}_{-0.3}$ & 
          (73 $^{+4}_{-4}$, 82 $^{+16}_{-15}$, 87$^{+7}_{-9})\times 10^{-3}$ & 0.25$^{+0.09}_{-0.06}$, 0.7$^{+0.3}_{-0.4}$\\
      Equivalent width (ms) & 34.1 $^{+2.6}_{-2.8}$ & 2.0 $^{+0.2}_{-0.2}$ & 8.6$^{+0.7}_{-0.8}$ & 0.32, 0.33, 0.35 & 1.24 $^{+0.13}_{-0.15}$\\
      Observed peak flux density, S$_{\rm peak}$ (Jy)& 5.19 & 15.75 & 10.19 & 161, 39, 89 & 19.23\\
      Fluence$^{\dagger}$ (Jy-ms) & $>$177 & $>$32 & $>$87 & $>$52, 13, 31 & $>$24\\
      
      \hline
      & & & \textbf{Model-dependent properties}  & & \\
      \hline
      DM$_{\rm MW}$$^{\ddagger}$ (\dm)& 45 & 70 & 89 & 39 & 58\\
      $\tau_{_{\rm MW}}$$^{\ddagger}$ ($\upmu$s) (at 835 MHz) & 0.31 & 0.76 & 1.58 & 0.21 & 0.49\\
      Max. inferred $z$     &  1.2       &   0.9   & 2.2 & 0.2 & 0.3 \\
      Max. comoving distance (Gpc)       &  3.8      &  3.1  &  5.5   & 0.8  & 1.2 \\
      Max. luminosity distance (Gpc)     &  8.1      & 5.9  & 17.4   & 0.9 & 1.6\\
      Max. isotropic energy (10$^{40}$ erg) & 21.2  & 2.4 & 31.9 & 0.4  & 0.2\\
      Peak luminosity (10$^{43}$ erg/s)     & 1.3   & 2.2 & 11.8  & 1.6 & 0.2\\
      \hline
      \multicolumn{5}{p{1.2\linewidth}}{$^{\ast}$\small{See Eq.\,\ref{eq:localisation_arc}}} \\
      \multicolumn{5}{p{1.2\linewidth}}{$^{\dagger}$\small{Corrected for the known position of the FRB within the primary beam pattern in the East-West direction, but uncorrected for the (unknown) FRB position in the north-south direction}}\\
      \multicolumn{5}{p{1.2\linewidth}}{$^{\ddagger}$\small{According to NE2001 model}}\\
    \end{tabular}
  \end{adjustbox}
\end{table*}

\section{FRB rate at 843 MHz}\label{sec:rates}

\begin{figure}
  \begin{center}
    \includegraphics[width=\linewidth]{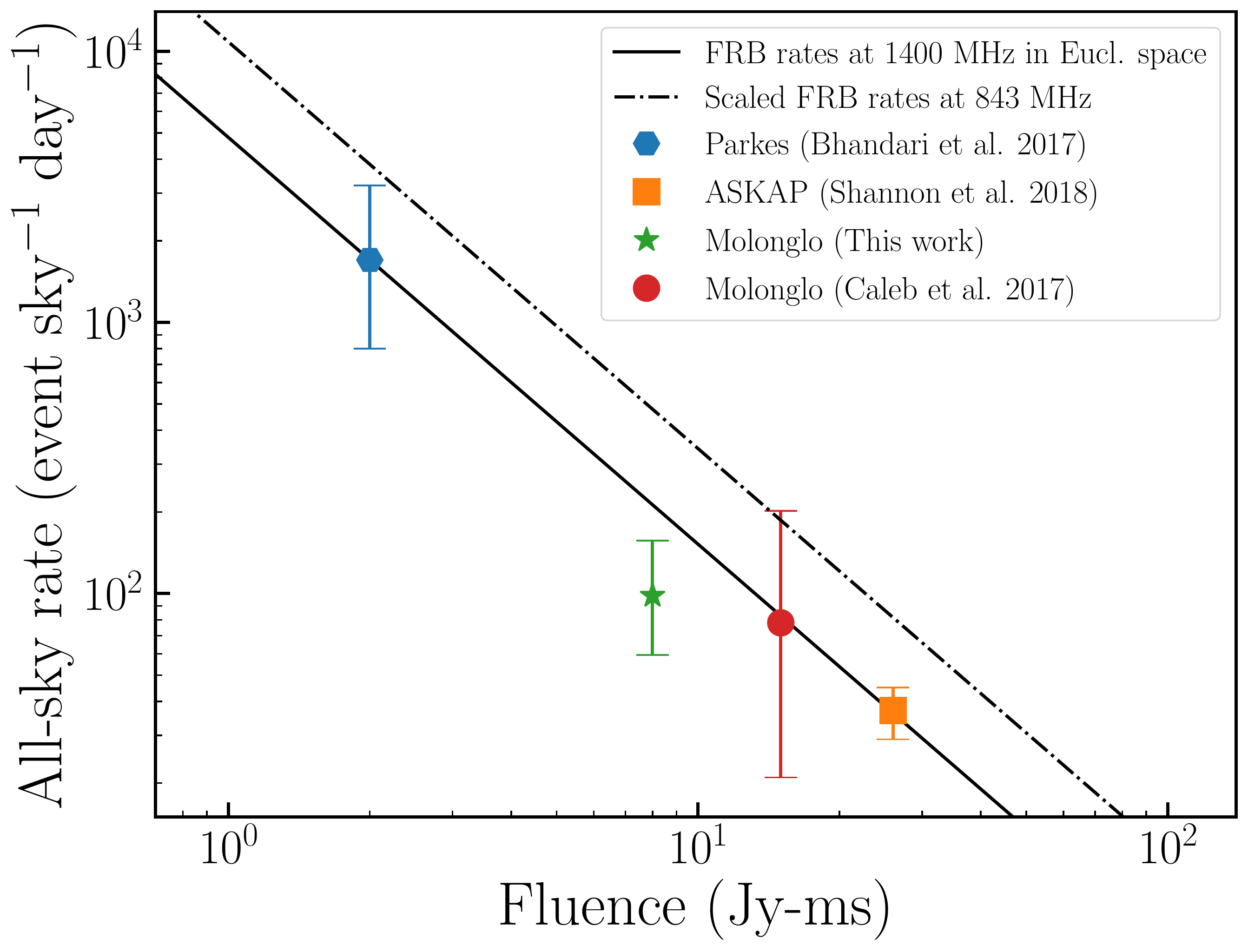}
    \caption{FRB rates in events/sky/day, shown as a function of fluence, at Parkes and ASKAP (at 1.4 GHz) and UTMOST (843 MHz). At UTMOST, we show the \citealt{Caleb_3frbs} event rate, based on the first 3 FRBs found (red circle) and the event rate reported here (green triangle) for 6 additional FRBs found in a more sensitive survey. The fluence limit was estimated as 11 Jy-ms by \citealt{Caleb_3frbs}: we have revised this to 15 Jy-ms, as our understanding of the flux calibration of UTMOST has improved markedly since the first 3 FRBs were found. The dotted line has a slope of $-1.5$ in this log-log plane, and represents the expected slope of the cumulative source counts for a Euclidean universe. It is a close match to the event rates seen in L-band (1.4 GHz) going from ASKAP to Parkes. The Assuming a flat spectral index for FRBs, the expected event rate at UTMOST is approximately 215 events/sky/day at a sensitivity of 8 Jy-ms. We obtain an rate of $98^{+59}_{-39}$ events/sky/day in the present survey, somewhat below the expected rate scaling from the 1.4 GHz rates. We also show (dashed line), the expected event rate at 843 MHz assuming FRBs have a mean spectral index of $-1.6$ \citep{Macquart2019_spec_idx}. At 8 Jy-ms sensitivity, we expect an event rate of $\approx 480$ events/sky/day. The event rate at UTMOST falls significantly below this value, inidcating that the mean spectral index of FRBs may not be this steep. }
    \label{rates}
  \end{center}
\end{figure}

The present survey ran from 2017 June 1 to 2018 December 31 commensally with the UTMOST pulsar timing/searching program (SMIRF --- Venkatraman Krishnan et al., in prep.). We estimate the total amount of time spent by UTMOST on sky during the  survey as 344 days. The survey yielded a total of 6 FRBs. 
Accounting for the efficiency of the detection pipeline (90 per cent, see \cref{subsec:validation}), we estimate the UTMOST FRB discovery rate as $\sim63$ 
days/event. This corresponds to a sky rate of $98^{+59}_{-39}$ events sky$^{-1}$ day$^{-1}$ above a fluence of 8 Jy-ms, where the quoted uncertainties represent 1-sigma Poissonian errors \citep{Gehrels1986}.

Fig.\,\ref{rates} shows UTMOST FRB sky rates (at 843 MHz) with our previous survey \citep{Caleb_3frbs} (red circle, based on 3 events) and for this survey (green triangle, based on 6 events). Note that, as a result of substantial improvements in our understanding of the flux calibration since the first 3 FRBs were found, we revise the fluence limit of the \citealt{Caleb_3frbs} survey from 11 to 15 Jy-ms as the authors overestimated the gain of the telescope. We also show the sky rates at 1.4 GHz measured at Parkes and ASKAP. The Parkes point (blue triangle) lies at 1700 events/sky/day down to 2 Jy-ms -- derived for the Parkes FRBs after taking fluence incompleteness into account \citep{Bhandari2018}. The ASKAP rate is also measured at 1.4 GHz and is 37 events/sky/day to a fluence of 26 Jy-ms as reported by \cite{Shannon2018}. 
The solid line shows the expected slope of the sky rate as a function of fluence for a Euclidean universe ($-1.5$ in 
this log-log plane). It appears to be a close match to the relative event rates going from bright events at ASKAP to weak events at 
Parkes. 
Assuming that FRBs have flat spectra, we would expect an event rate at UTMOST, 
interpolating between Parkes and ASKAP, of approximately 215 events/sky/day at a sensitivity of 8 Jy-ms. The observed rate however, is 
$98^{+59}_{-39}$ events/sky/day in the present survey. This observed rate at 843 MHz is therefore approximately 2-$\sigma$ below the expected rate scaling from ASKAP and Parkes for the simple model of Euclidean 
counts and flat spectrum sources. We also show (dashed line) the expected event rate at 843 MHz assuming FRBs have a mean spectral index 
of $-1.6^{+0.3}_{-0.2} $ \citep{Macquart2019_spec_idx}. At 8 Jy-ms sensitivity, we expect an event rate of $\sim 480$ events/sky/day. The UTMOST event rate falls significantly ($\sim 7$-$\sigma$) below this value, arguing against such a steep spectral index. 

Given the lower than expected rate at 843 MHz suggests that the spectra of FRBs may turn over at about 1 GHz. This is consistent with a number of recent studies. Firstly, 6 ASKAP FRBs were observed simultaneously with the Murchison Wide Field array (MWA) but yielded only upper limits on their fluences at 170-200 MHz, indicating that the spectral index of FRBs is no steeper than $\alpha \approx -1$ \citep{Sokolowski2018}. Secondly, and more significantly, the non-detection of FRBs in an 84-day survey made at the Green Bank Telescope (GBT) \citep{Chawla2017} at 300-400 MHz to a sensitivity of 0.6 Jy-ms (for 5-ms events), places an upper limit on the spectral index of FRBs of $\alpha > -0.3$. \cite{Ravi&Loeb2019} discuss these results in detail and propose a number of mechanisms to explain why the spectral energy distribution of FRBs would turn over below $\approx 1$ GHz. The UTMOST results reported here are consistent with these proposals.

\begin{figure}
  \begin{center}
    \includegraphics[width=\columnwidth]{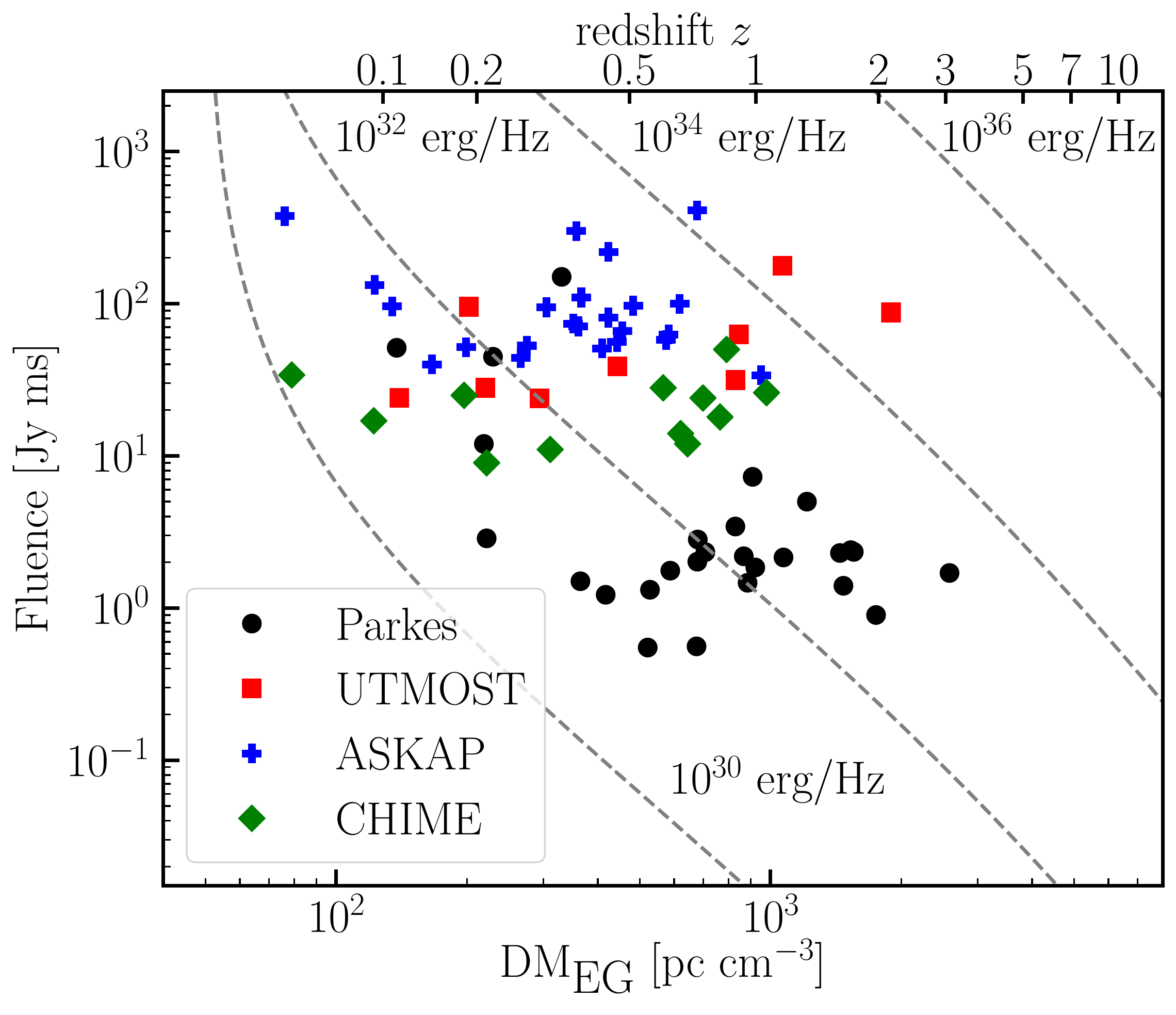}
    \centering
    \caption{The fluence of FRBs found at Parkes (black), ASKAP (blue) and UTMOST (red) are shown as a function of their extragalactic 
    dispersion measures, DM$_\textrm{EG}$, following Shannon et al. (2018). 
    Dashed lines show the fluence evolution with DM for constant spectral energy density sources, due to cosmological effects. The 
    redshift scale corresponding to the DM axis is shown at the top of the plot
    We assume FRBs have flat spectra and that 83 percent of baryons are in the IGM which is fully ionised at all $z$ and is composed of 
    0.75 H and 0.25 He by mass fraction (Zhang 2018). 
    }
    \label{figure:shannon_plot}
  \end{center}
\end{figure}

In Fig.\,\ref{figure:shannon_plot}, we show FRB fluences versus extragalactic DM for our sample of 9 FRBs (red squares) at 830--850 MHz, 
23 ASKAP FRBs (blue crosses) at 1.2--1.6 GHz, 13 CHIME FRBs (green diamonds) at 400--800 MHz and 19 Parkes FRBs (black circles) at 
1.2--1.6 GHz. Lines of constant energy density are shown for a standard cosmology (see figure caption) with 
the important assumption that FRB spectra are flat. We adopt this assumption to simplify the comparison of FRBs found in surveys with 
very different frequency coverage, but note that
FRBs might not have flat spectra, as discussed above. As has been 
argued by \cite{Shannon2018}, the trend to lower fluence with increasing DM argues for FRBs indeed being at cosmological distances 
(upper scale of plot). It is clear that FRBs span a wide range of intrinsic energies (of order 2 decades) at a given DM, 
indicating their intrinsic luminosity function is broad. Our results show that this trend still
holds for FRBs discovered with different instruments operating at different wavelengths, channel widths, and 
integration times.



\section{FRB follow-up}\label{sec:frb_followup}

\subsection{Radio Follow-up}
As part of the dynamic scheduling of observations by SMIRF, the fields of our own FRBs and a selection of those found in the ASKAP/CRAFT project were regularly re-observed to search for FRB repetition. The FRB fields searched and the total observing time for each since deployment of the SMIRF scheduler are listed in Table \ref{table:followup}. A total of 120 hours of follow-up at UTMOST was performed for 23 FRB fields. Typically, observations had a duration of the transit time of the field centre across the FWHP of the primary beam (4 degrees) and, depending on the declination of the FRB, is $\sim$20 minutes. No FRBs were seen to repeat
during the follow-up program down to a S/N of 9.  

Motivated by the resemblance --- in temporal and spectral structure --- of FRB181017 to the repeating FRB121102 \citep{Hessels2018}, we conducted a follow-up campaign to search for repeating bursts using more sensitive facilities: the Effelsberg radio telescope and the upgraded Giant Metrewave Radio Telescope (uGMRT).

\textbf{Effelsberg:} 
Data were obtained on UTC 2018 October 25 and UTC 2018 November 05 using the 7-beam feed
array and the high time resolution (54 microseconds) Pulsar Fast Fourier Transform
Spectrometer (PFFTS) backend in pulsar search mode (Barr et al. 2013). The data
was centered at a frequency of 1.36 GHz with a bandwidth of 300 MHz divided over
512 channels. The receiver was rotated such that 3 of the 7 beams were
aligned along the uncertainty arc of the FRB. The localisation arc was tiled with 11
partially overlapping pointings (33 beams of 10’ each) along its North-South extend of
~2.8 degrees. We searched for pulses in these 3 beams using \textsc{heimdall} over a range of 30 pc cm$^{-3}$, centered on the DM of the FRB, and pulse widths in the range 54 $\upmu$s to 55 ms, down to a S/N of 7. We required that candidate events appear in 1 beam of the instrument only. This corresponds to a search sensitivity of 0.2 Jy-ms for a 1 ms pulse. We found no repeat bursts of the FRB.

\textbf{uGMRT:} Observations of FRB181017 were made on UTC 2018 November 17, 2018 November 27 and 2018 November 29 with the incoherent uGMRT array in 
band-4 (550-850 MHz). Data were recorded at 327.68 $\upmu$s with 8192 channels over the band to ensure that the dispersion smearing 
within a channel is comparable to the time-resolution at the DM of the FRB. As the FWHM of the uGMRT beam in this band is 
$\sim37\arcmin$, the uncertainty in the FRB declination was tesselated into a strip of 10 individual overlapping pointings at the 
nominal RA. The data were searched offline using the \textsc{heimdall} single pulse search software for pulses with S/N $\geq 6$, DMs in 
the range $220 \leq \mathrm{DM} \leq 260$ pc cm$^{-3}$ and widths $\leq 100$ ms. RFI mitigation was performed using the \texttt{clfd}\footnote{\href{https://github.com/v-morello/clfd}{https://github.com/v-morello/clfd}} 
package described in \cite{Morello2018}. We did not find any repeat pulses from the FRB in a total of 8.3 hours spent on source.

\subsection{Optical Follow-up}

For 3 of the FRBs reported here (FRB170922 was discovered two weeks after data recording, 
and FRB181016 was discovered during the Australian daytime), 
a search for possible optical afterglow was conducted using the SkyMapper
telescope \citep{Keller2007}. 
We established an automated system that allows scheduling of an FRB field to 
be triggered via email. The shortest time from FRB trigger to observations has been $\sim$2 hours but is 
typically the following night or nights, contingent on weather and field location relative to the Sun and Moon. 

FRB181017: no useful science images were produced due to bad weather conditions on site, 
a 70\% illuminated moon and its close proximity ($\sim 15$ degrees)
to the centre of the FRB localisation arc.

FRB180528 \& FRB181228: images were taken in the $r$ and $i$ bands 
for which the photometric depths for a 100 second exposure are 
$i=19.17$, $r=19.54$ (FRB180528) and $i=20.7$, $r=21.7$ (FRB181228)
at the 95\% upper limit (SkyMapper Transient Survey Pipeline, \citealt{Scalzo2017}). 

The follow-up fields were centred on the most likely FRB coordinate as reported in our Astronomer's telegrams along 
with fields to the north and south to cover the 1-sigma uncertainty in the localisation 
arcs for FRBs detected with the current operation mode at UTMOST (i.e. 4.8 degrees). 
Observations consist of multiple images centered on the FRB 
most likely positions, with slight pointing offsets, 
followed by imaging of the 1-sigma regions. The localisation arc of each FRB 
was searched for optical transients with reference to existing images from SkyMapper's database, or with reference to 
images taken on subsequent nights. We found no optical transients that could be associated with our FRB events.

\begin{table}
  \centering
  \begin{threeparttable}
    \caption{FRB field follow-up campaign with UTMOST. No repeat pulses were found for any FRB.}
    \label{table:followup}
    \begin{tabular}{l c c c}
      \toprule
      \toprule
      FRB name & Total time (hours) & Discovery & Reference$^{1}$\\
      \toprule
      FRB160317 & 6.3 & UTMOST & [1] \\
      FRB160410 & 2.0 & UTMOST & [1] \\
      FRB160608 & 3.9 & UTMOST & [1] \\
      FRB170107 & 2.0 & ASKAP & [2] \\
      FRB170416 & 5.8 & ASKAP & [3] \\
      FRB170428 & 5.8 & ASKAP & [3] \\
      FRB170707 & 8.2 & ASKAP & [3] \\
      FRB170712 & 10.2 & ASKAP & [3] \\
      FRB170827 & 29.8 & UTMOST & [4] \\
      FRB170906 & 5.1 & ASKAP & [3] \\
      FRB170922 & 10.5 & UTMOST & This work \\
      FRB171003 & 1.9 & ASKAP & [3] \\
      FRB171004 & 2.4 & ASKAP & [3] \\
      FRB171019 & 5.9 & ASKAP & [3] \\
      FRB171020 & 5.8 & ASKAP & [3] \\
      FRB171116 & 3.5 & ASKAP & [3] \\
      FRB171213 & 4.4 & ASKAP & [3] \\
      FRB171216 & 4.2 & ASKAP & [3] \\
      FRB180110 & 4.2 & ASKAP & [3] \\
      FRB180119 & 4.2 & ASKAP & [3] \\
      FRB180309 & 0.5 & Parkes & [5] \\
      FRB180528 & 6.0 & UTMOST & This work \\
      FRB181016 & 2.7 & UTMOST & This work \\
      FRB181017 & 8.1 & UTMOST & This work \\
      \bottomrule
    \end{tabular}
    \begin{tablenotes}
      \item $^{1}$ [1] \cite{Caleb_3frbs},  [2] \cite{Bannister2017},  [3] \cite{Shannon2018},  [4] \cite{Farah2018}, [5] \cite{oslowski2018_atel_superbright}.
    \end{tablenotes}
  \end{threeparttable}
\end{table}

\subsection{FRB181228 follow-up}

An astronomer's telegram for FRB181228 \citep{Farah_FRB181228_atel} was issued within 2 hours of the event, and there has been considerable 
follow-up by external parties, attesting to the efficacy of early triggering. No counterparts 
have been found. An optical transient was found with MASTER PN \citep{2018ATel12338} in a region close to the 
localisation arc. This was determined to be a type Ia supernova after spectroscopy was obtained with the Southern African Large Telescope \citep{2018ATel12343}. 
They report the source is likely to be 10 days post-maximum and hosted in the galaxy LEDA 499631, with a redshift in 
the range 0.025 to 0.031. The maximum DM inferred redshift of FRB181228 is 0.3. It 
is thus unlikely that the type Ia supernova is associated with the FRB. 
X-ray data from Astrosat CZTI was also searched for an 
associated transient in a 20 second window, with no counterpart found \citep{2018ATEL12370}. 

\section{Conclusions}\label{sec:conclusions}


We have presented the results of the latest FRB survey conducted with the Molonglo radio telescope,
using a newly-implemented live machine-learning based FRB detection system. 
We accumulated a total of 344 days on sky 
searching for FRBs in real time, discovering 6 FRBs.

We demonstrated the importance of the real-time detection of FRBs, as evidenced by the 
discovery of high time and frequency structure in FRB pulse profiles
resulting from the capture of the raw data --- particularly for our higher S/N events. 
This has allowed us to probe the properties of some of the narrowest and least scattered FRBs to date. 
The temporal profile of FRB181017 shows 3 peaks, with the middle component not centred in 
time. This argues against a source of underlying periodicity 
on the $\sim 1$ ms timescales. The FRB dynamic spectrum is similar to our other bright event (FRB170827), 
as well as to the first repeating FRB (FRB121102), potentially linking repeating and non-repeating FRBs. 
The frequency structure across the multi-peaked profile FRBs argues for an origin associated with 
the propagation in the host galaxy or the IGM, rather than arising at the source. 
Moreover, given the triple-peak temporal structure of this FRB, we 
rule out a lensing scenario by finding no evidence that 
the voltage data of the leading and trailing peaks are correlated. 
We encourage the application of this technique to multi-component FRBs soon 
to be found with new generation telescopes such as CHIME, MeerKAT, ASKAP and UTMOST-2D.

We derive an event rate of 
$98^{+59}_{-39}$ events/sky/day at a fluence limit of 8 Jy-ms at 843 MHz. 
This rate is somewhat below expectation, scaling from the FRB rates found at Parkes and ASKAP, both of which operate 
at 1.4 GHz, and assuming that the average spectral energy distribution of FRBs is flat. Our results do not agree with 
the steep negative spectral index estimates for mean FRB spectra of $\approx -1.6 \pm 0.2$ 
\citep{Macquart2019_spec_idx}, and may indicate that the spectra of FRBs turnover at around 1 GHz, as has been recently 
suggested by \cite{Ravi&Loeb2019}. The CHIME collaboration has reported 
13 FRBs in the range 400-800 MHz, and estimate a lower limit on the sky rate of 
300 event sky$^{-1}$day$^{-1}$ to a flux density of 1 (ms$/\Delta t)^{1/2}$ Jy. Their very high discovery rate should
allow the question of a turnover in the spectral energy density of FRBs to be probed 
in the near future.

We are currently outfitting the NS arm of the telescope for the UTMOST-2D project, which will provide 
localisations of FRBs from single detections with arcsecond precision. The highly 
effective machine-learning FRB live detection pipeline reported here will be used to 
trigger full data retention of single pulse events, as a major part of our hunt for 
FRB hosts.

\vskip 1.0 truecm
\section*{Acknowledgements}
We thank Michael Kramer for valuable discussions on FRB scattering tail fitting. We thank Laura Spitler for assistance
with the Effelsberg data analysis. The Molonglo Observatory is owned and operated by the University of Sydney, with support from the School of 
Physics and the University. The UTMOST project is also supported by the Swinburne University of Technology. We acknowledge the Australian Research Council grants CE110001020 (CAASTRO) and the Laureate Fellowship FL150100148.
ATD is supported by an ARC Future Fellowship grant FT150100415. 
MC and BWS acknowledge funding from the European Research Council (ERC) under the European Union’s Horizon 2020 
research and innovation programme (grant agreement no. 694745).
The national facility capability for SkyMapper has been funded through 
the Australian Research Council LIEF grant LE130100104, awarded
to the University of Sydney, the Australian National University,
Swinburne University of Technology, the University of Queensland,
the University of Western Australia, the University of Melbourne,
Curtin University of Technology, Monash University, and the Australian Astronomical Observatory.
SkyMapper is owned and operated by The Australian National University's Research School of Astronomy and Astrophysics.
The GMRT is run by the National Centre for Radio Astrophysics of the Tata Institute of Fundamental 
Research, India. We acknowledge 
support of GMRT telescope operators for the observations.
This research has made use of NASA's Astrophysics Data System.

\bibliographystyle{mnras}
\bibliography{bibliography}

\bsp	
\label{lastpage}
\end{document}